\documentclass[twocolumn]{aastex62}

\graphicspath{{./}{figures/}}

\usepackage{amsmath}
\received{}
\revised{}
\accepted{}
\submitjournal{ApJ}
\usepackage{graphicx,animate}

\shorttitle{REQUIEM-2D: A diversity of formation pathways in massive quiescent galaxies}
\shortauthors{M. Akhshik, K. Whitaker, J. Leja, et. al.}

\begin{document}

\title{REQUIEM-2D: A diversity of formation pathways in a sample of spatially-resolved massive quiescent galaxies at $z\sim2$}

\correspondingauthor{Mohammad Akhshik}
\email{mohammad.akhshik@uconn.edu}

\author[0000-0002-3240-7660]{Mohammad Akhshik}
\affil{Department of Physics, University of Connecticut,
Storrs, CT 06269, USA}

\author[0000-0001-7160-3632]{Katherine E. Whitaker}
\affil{Department of Astronomy, University of Massachusetts, Amherst, MA 01003, USA}
\affil{Cosmic Dawn Center (DAWN), Denmark}

\author[0000-0001-6755-1315]{Joel Leja}
\affil{Department of Astronomy \& Astrophysics, The Pennsylvania State University, University Park, PA 16802, USA}
\affil{Institute for Computational \& Data Sciences, The Pennsylvania State University, University Park, PA, USA}
\affil{Institute for Gravitation and the Cosmos, The Pennsylvania State University, University Park, PA 16802, USA}

\author[0000-0001-5492-1049]{Johan Richard}
\affil{Univ Lyon, Univ Lyon1, Ens de Lyon, CNRS, Centre de Recherche Astrophysique de Lyon UMR5574, F-69230, Saint-Genis-Laval, France}

\author[0000-0003-3256-5615]{Justin~S.~Spilker}
\affiliation{Department of Physics and Astronomy and George P. and Cynthia Woods Mitchell Institute for Fundamental Physics and Astronomy, Texas A\&M University, 4242 TAMU, College Station, TX 77843-4242, US}

\author[0000-0002-8442-3128]{Mimi Song}
\affil{Department of Astronomy, University of Massachusetts, Amherst, MA 01003, USA}

\author[0000-0003-2680-005X]{Gabriel Brammer}
\affil{Cosmic Dawn Center (DAWN), Denmark}
\affil{Niels Bohr Institute, University of Copenhagen, Jagtvej 128, DK-2200 Copenhagen, Denmark}

\author[0000-0001-5063-8254]{Rachel Bezanson}
\affil{Department of Physics and Astronomy, University of Pittsburgh, Pittsburgh, PA 15260, USA}

\author[0000-0001-8249-2739]{Harald Ebeling}
\affil{Institute for Astronomy University of Hawaii, 2680 Woodlawn Drive Honolulu, HI 96822, USA}

\author[0000-0002-9656-1800]{Anna R. Gallazzi}
\affil{INAF-Osservatorio Astrofisico di Arcetri, Largo Enrico Fermi 5, 50125 Firenze, Italy}

\author[0000-0003-3266-2001]{Guillaume Mahler}
\affil{Institute for Computational Cosmology, Durham University, South Road, Durham DH1 3LE, UK}
\affil{Centre for Extragalactic Astronomy, Durham University, South Road, Durham DH1 3LE, UK}

\author[0000-0002-8530-9765]{Lamiya A. Mowla}
\affil{Dunlap Institute for Astronomy and Astrophysics, University of Toronto, 50 St George Street, Toronto ON, M5S 3H4, Canada}

\author[0000-0002-7524-374X]{Erica J. Nelson}
\affil{Department of Astrophysical and Planetary Sciences, 391 UCB, University of Colorado, Boulder, CO 80309-0391, USA}

\author[0000-0003-4196-0617]{Camilla Pacifici}
\affil{Space Telescope Science Institute, 3700 San Martin Drive, Baltimore MD 21218, USA}

\author[0000-0002-7559-0864]{Keren Sharon}
\affil{Department of Astronomy, University of Michigan, 1085 South University Ave, Ann Arbor, MI 48109, USA}

\author[0000-0003-3631-7176]{Sune Toft}
\affil{Cosmic Dawn Center (DAWN), Denmark}
\affil{Niels Bohr Institute, University of Copenhagen, Jagtvej 128, DK-2200 Copenhagen, Denmark}

\author[0000-0003-2919-7495]{Christina C. Williams}
\affil{Steward Observatory, University of Arizona, 933 North Cherry Avenue, Tucson, AZ 85721, USA}

\author{Lillian Wright}
\affil{Department of Astronomy, University of Massachusetts, Amherst, MA 01003, USA}

\author[0000-0002-9842-6354]{Johannes Zabl}
\affil{Institute for Computational Astrophysics and Department of Astronomy \& Physics, Saint Mary’s University, 923 Robie Street, Halifax, Nova Scotia, B3H 3C3, Canada}

\begin{abstract}

REQUIEM-2D (REsolving QUIEscent Magnified galaxies with 2D grism spectroscopy) is comprised of a sample of 8 massive ($\log M_*/M_\odot > 10.6$) strongly lensed quiescent galaxies at $z\sim2$. REQUIEM-2D combines the natural magnification from strong gravitational lensing with the high spatial-resolution grism spectroscopy of  \emph{Hubble Space Telescope} through a spectrophotometric fit to study spatially resolved stellar populations. We show that quiescent galaxies in the REQUIEM-2D survey have diverse formation histories manifesting as a gradient in stellar ages, including examples of (1) a younger central region supporting outside-in formation, (2) flat age gradients that show evidence for both spatially-uniform early formation or inside-out quenching, and (3) regions at a fixed radial distance having different ages (such asymmetries cannot be recovered when averaging  stellar population measurements azimuthally). The typical dust attenuation curve for the REQUIEM-2D galaxies is constrained to be steeper than Calzetti's law in the UV and generally consistent with $A_V<1$.  Combined together and accounting for the different physical radial distances and formation time-scales, we find that the REQUIEM-2D galaxies that formed earlier in the universe exhibit slow and uniform growth in their inner core, whereas the galaxies that formed later have rapid inner growth in their inner core with younger ages relative to the outskirts.  These results challenge the currently accepted paradigm of how massive quiescent galaxies form, where the earliest galaxies are thought to form most rapidly. Significantly larger samples close to the epoch of formation with similar data quality and higher spectral resolution are required to validate this finding.
\end{abstract}


\section{Introduction} \label{sec:intro}
\begin{figure}[!t]
\centering
\includegraphics[width=\columnwidth]{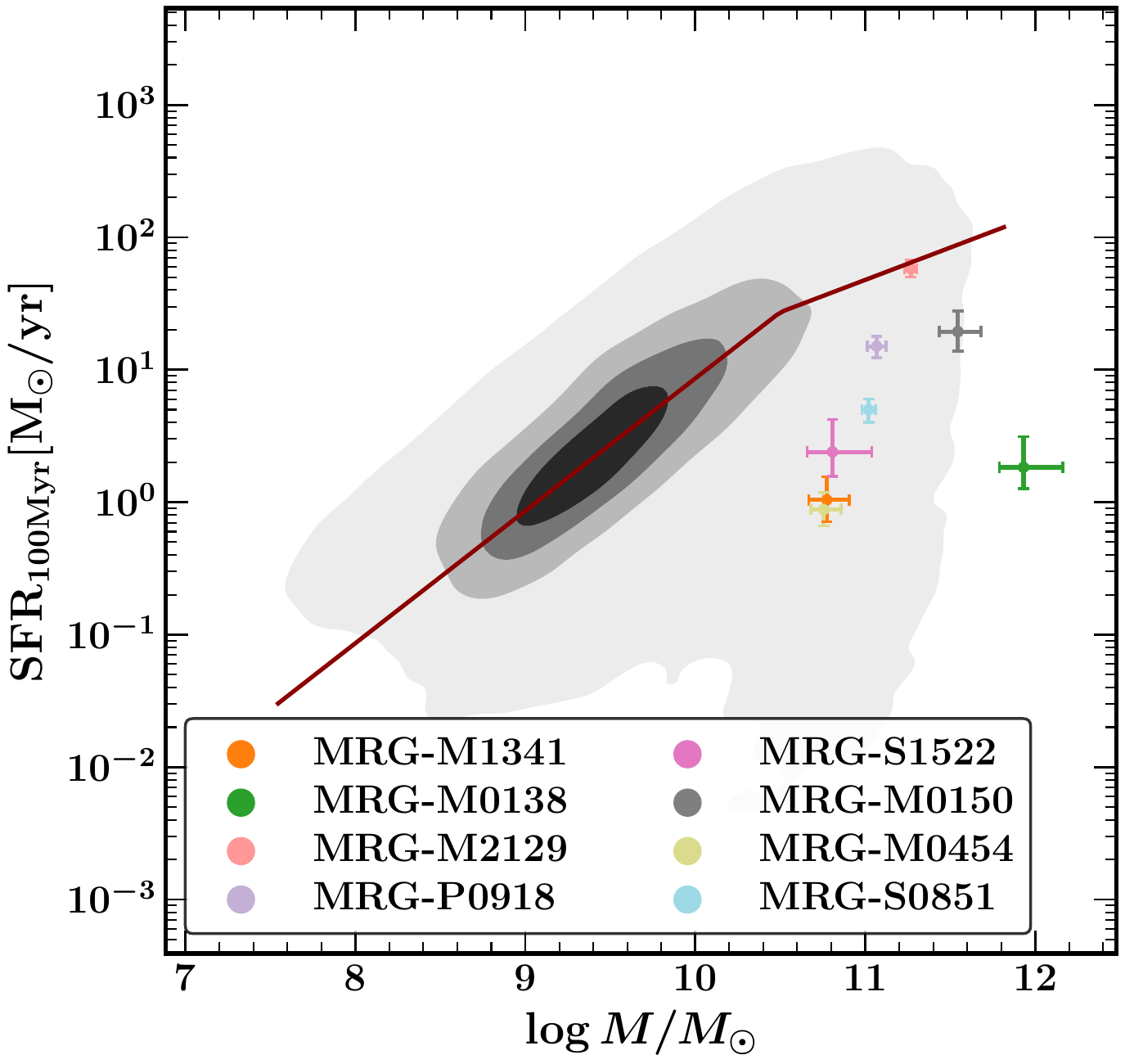}
\caption{SFMS from the 3D-HST survey \citep{brammer2012a,momcheva2016}. The stellar populations' parameters are constrained using \texttt{Prospector-$\alpha$} \citep{johnson2019, leja2017, leja2020}. The red line shows the best fit to the SFMS at $z=1$-$3$ following \citet{leja2021,whitaker2014}. The REQUIEM-2D galaxies are shown with the circles. For the REQUIEM-2D galaxies, the stellar masses are the total stellar mass formed in all spatial bins and the SFR is sum of the SFR of individual bins (see Sections \ref{sec:data_red}-\ref{sec:resolved-results}). MRG-M2129 is on the SFMS because it has a dust-obscured star-forming component (see Section \ref{sec:discussion}). Both parameters are corrected assuming a linear gravitational magnification. \label{fig:sfms}}
\end{figure}

One of the key empirical findings in observational extragalactic astronomy is the existence of a star-formation main sequence (SFMS) manifesting as a population of galaxies with a relatively tight correlation between their star-formation rate (SFR) and stellar mass \citep[e.g.][see also Figure \ref{fig:sfms}]{elbaz2007, daddi2007, noeske2007, whitaker2012, whitaker2014, speagle2014}. It is also shown observationally that there is a population of massive galaxies with significantly lower SFR than what is expected from the SFMS, known as quiescent galaxies \citep[e.g.][]{brammer2009,kriek06,muzzin2013}. Understanding the existence of the SFMS \citep[e.g.,][]{kelson2014} and its outliers is one of the major challenges of extragalactic astronomy.

Spectrophotometric observations, that include UV to far IR data, are extensively used to study quiescent galaxies at $z\gtrsim 1$ \citep[e.g.,][]{geier2013, hill2016,  belli2017, toft2017, abramson2018, ebeling2018, newman2018, morishita2018, morishita2019, belli2019, belli2021,  estrada2019, estrada-carpenter2020,  jafariyazani2020, setton2020, man2021, tachella2021,   whitaker2013, whitaker2021}. These studies imply a diversity among the progenitors, including compact galaxies that form stars centrally \citep[e.g.,][]{williams2014} and rotationally supported larger star-forming galaxies \citep[e.g.,][]{wisnioski2015}. Quiescent galaxies are thought to form via either ``slow'' or ``fast'' channels, leading to different colors, formation time-scales and the ages of the stellar populations \citep[e.g.,][]{belli2019}. The cold gas content of quiescent galaxies has also been studied to some extent, revealing a medley of values for the inferred depletion timescales and gas to stellar-mass ratios \citep[e.g.,][]{belli2021, bezanson2019, man2021, spilker2018, suess2017, whitaker2021, williams2021}. These early studies find cases that support both fast depletion of the cold gas and/or reduced star-formation efficiency  associated with the quiescent population.

There is an equally diverse set of potential mechanisms for explaining the existence of quiescent galaxies in cosmological simulations \citep[e.g.,][]{dekel2006, Tacchella2015, wellons2015, zolotov2015, tacchella2016}, motivating observational studies in order to constrain the imprints of the corresponding physical processes on their stellar populations. Specifically, constraining the spatial variation of stellar age within the galaxies can shed light on their formation pathways \citep[e.g.,][]{wellons2015,tacchella2016}. However, as a stellar population gets older, the variations in the age-sensitive spectrophotometric features decrease, making it very difficult to detect age gradients for an older age baseline. This limits the ideal candidates to  ``recently''\footnote{roughly defined as galaxies that have been actively forming stars $\lesssim1$ Gyr before observing.} quenched quiescent galaxies, which peak in number density at $z\gtrsim 1$ \cite[e.g.,][]{whitaker2012, cassata2013}.

We present the REQUIEM-2D survey in this paper, studying the stellar populations of quiescent galaxies that cover a range of global ages, stellar masses and star-formation rates \citep[see][and Section \ref{sec:data_red}]{akhshik2020}. This survey is designed to use a specific feature of the \emph{HST} Wide Field Camera 3 (WFC3) grism, its ability to preserve the spatial information perpendicular to the dispersion angle, to study spatially-resolved stellar populations. When combining the high spatial-resolution of \emph{HST} with a natural boost from strong gravitational lensing, the REQUIEM-2D survey provides a unique opportunity to constrain the spatially-resolved ages and star-formation histories of the inner cores of 8 massive quiescent lensed galaxies in order to gain further insights into their formation pathways \citep[see also][for a similar study]{abramson2018}. 

Herein, we present the analyses of the 7 targets from the REQUIEM-2D survey; a detailed study of the REQUIEM-2D's pilot target, MRG-S0851 is presented elsewhere \citep{akhshik2020,akhshik2021, caliendo2021}. We summarize the data reduction in Section \ref{sec:data_red}, and present the analysis in Sections \ref{sec:resolved-analysis} and \ref{sec:resolved-results}, referring the reader to \citet{akhshik2020} for a more in depth discussion of the adopted methodology. In Sections \ref{sec:discussion} and \ref{sec:summary} we discuss the implications of our results. The details of the previously unpublished lensing model for MRG-P0918 are presented in Appendix \ref{sec:lensing}.

In this paper, we assume a standard simplified $\mathrm{\Lambda CDM}$ cosmology with $\Omega _{M}=0.3$, $\Omega _{\Lambda}=0.7$ and $H_0 = 70 \, \mathrm{km}\mathrm{s}^{-1}\mathrm{Mpc}^{-1}$ and the \citet{chabrier2003} initial mass function. All magnitudes are reported in the AB system.

\begin{figure*}[!th]
\centering
\begin{tabular}{ccc}
\includegraphics[width=0.32\textwidth]{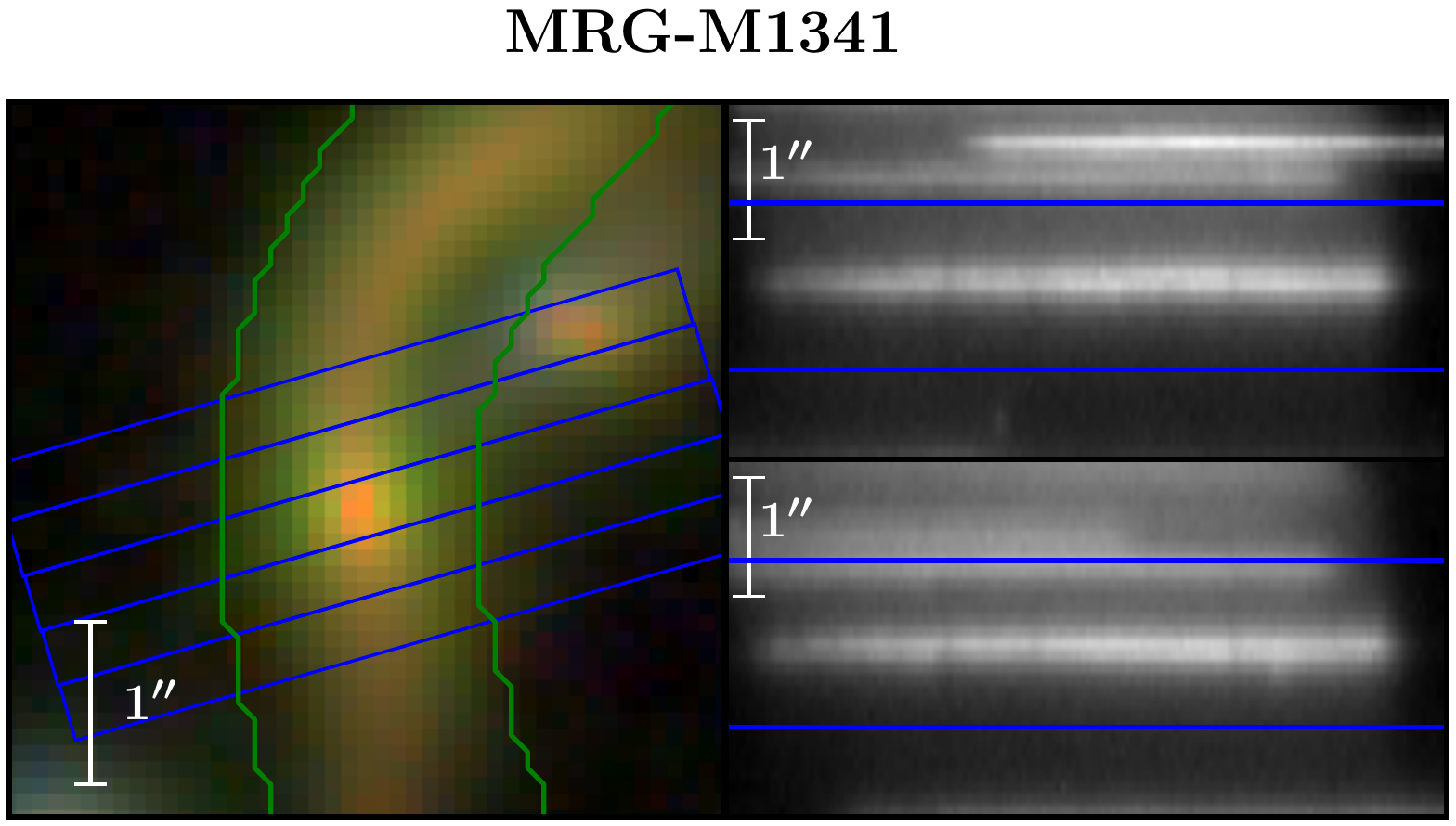}
&
\includegraphics[width=0.32\textwidth]{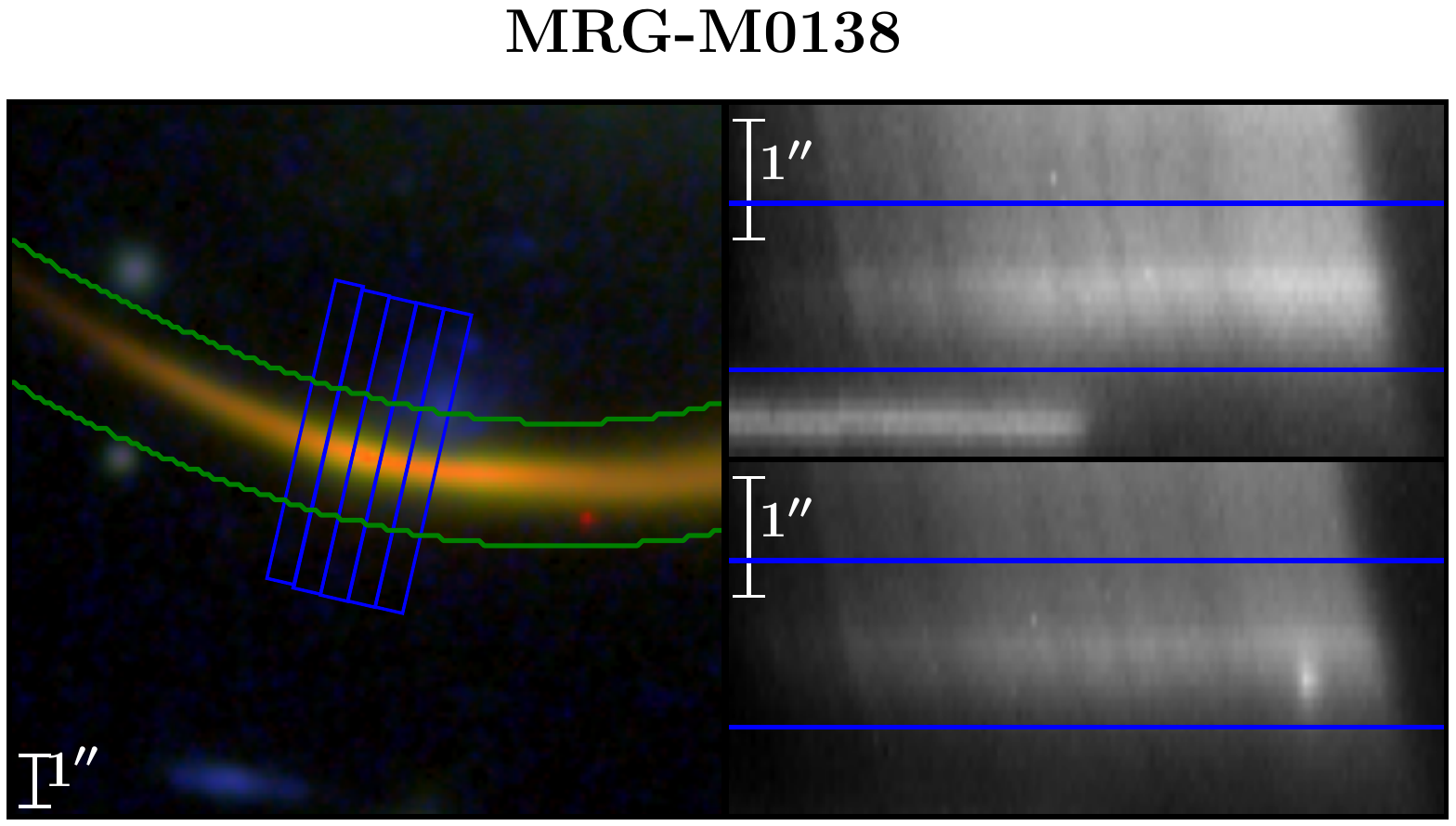}
&
\includegraphics[width=0.32\textwidth]{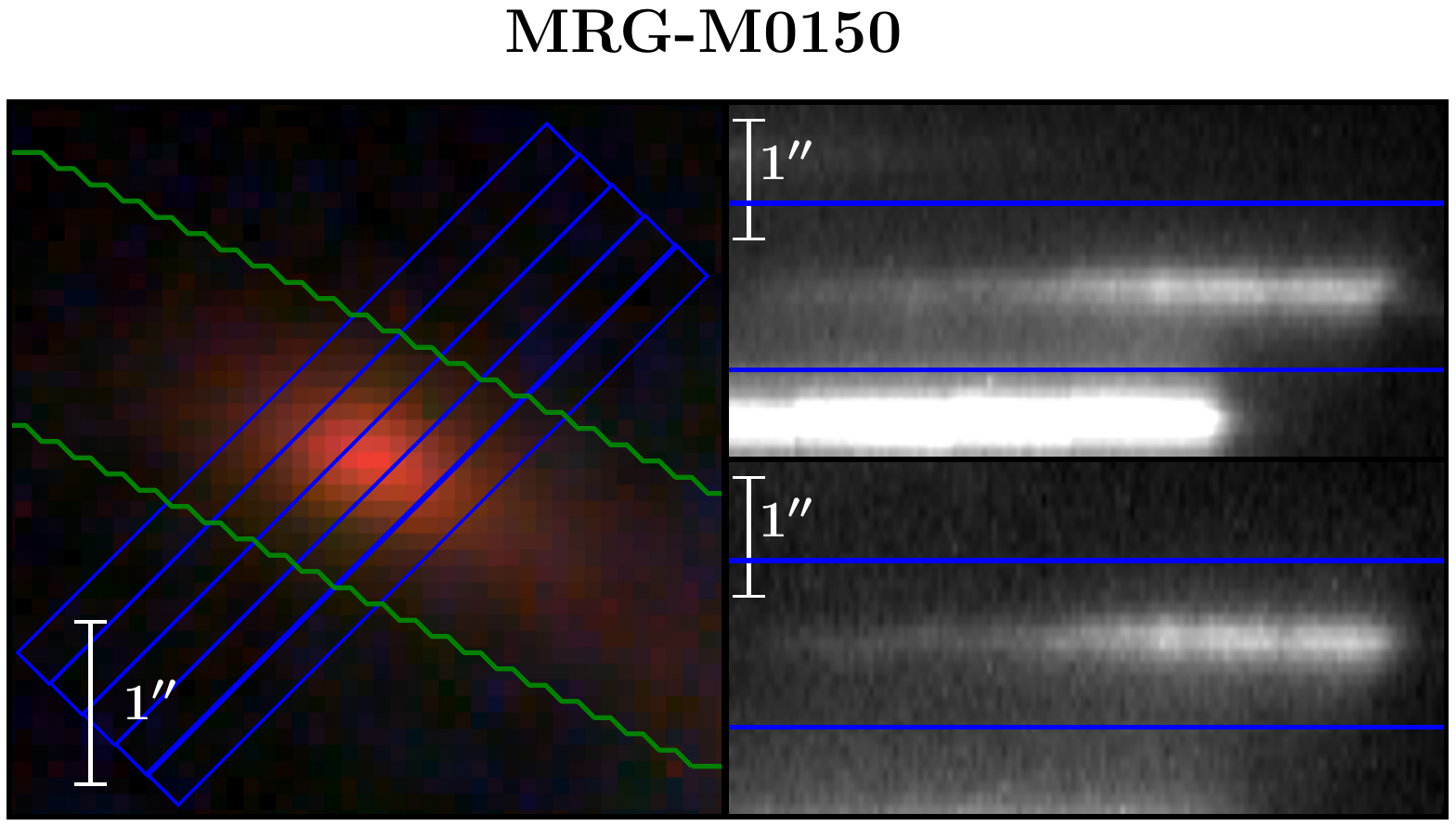}
\\
\includegraphics[width=0.32\textwidth]{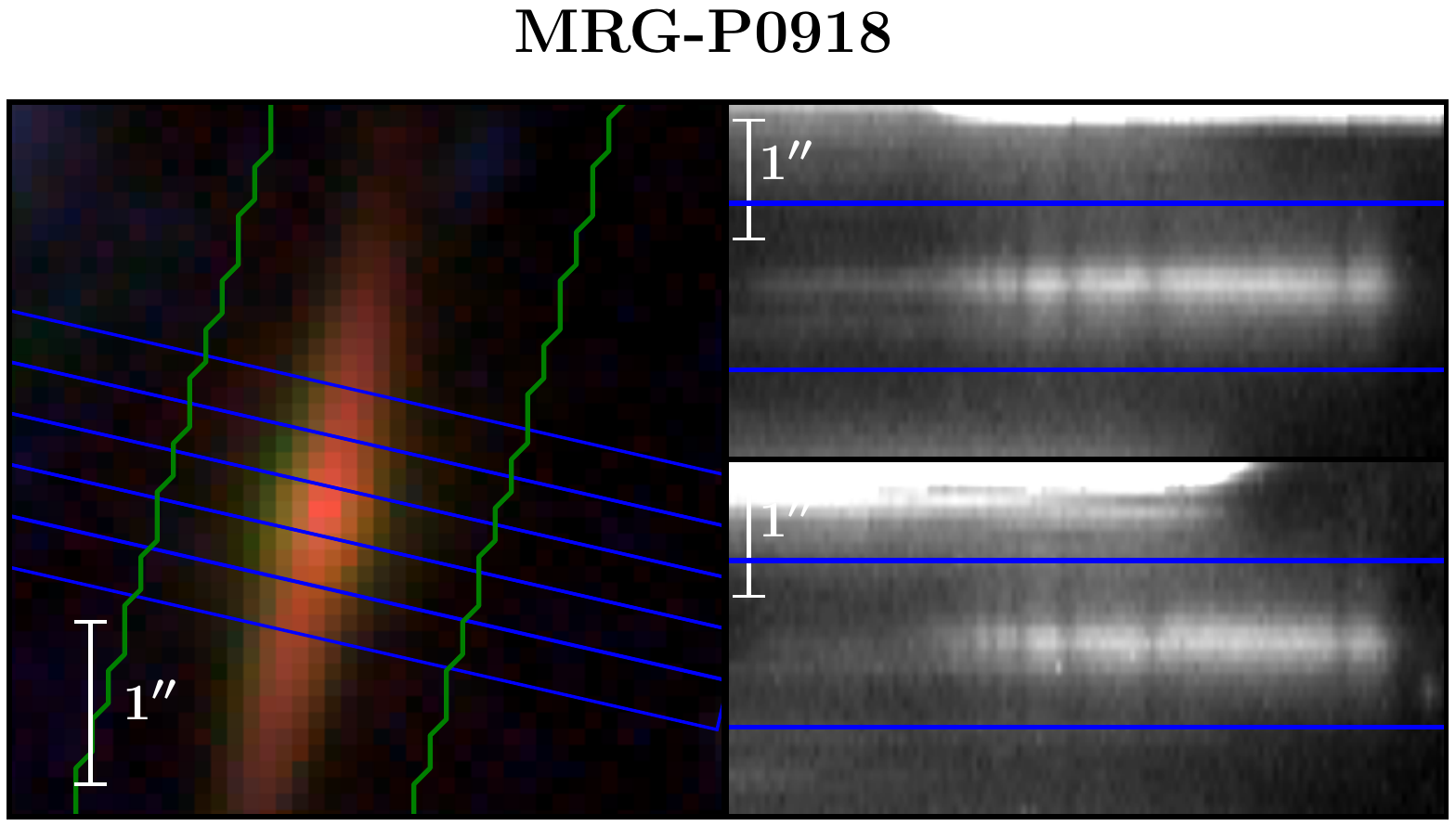}
&
\includegraphics[width=0.32\textwidth]{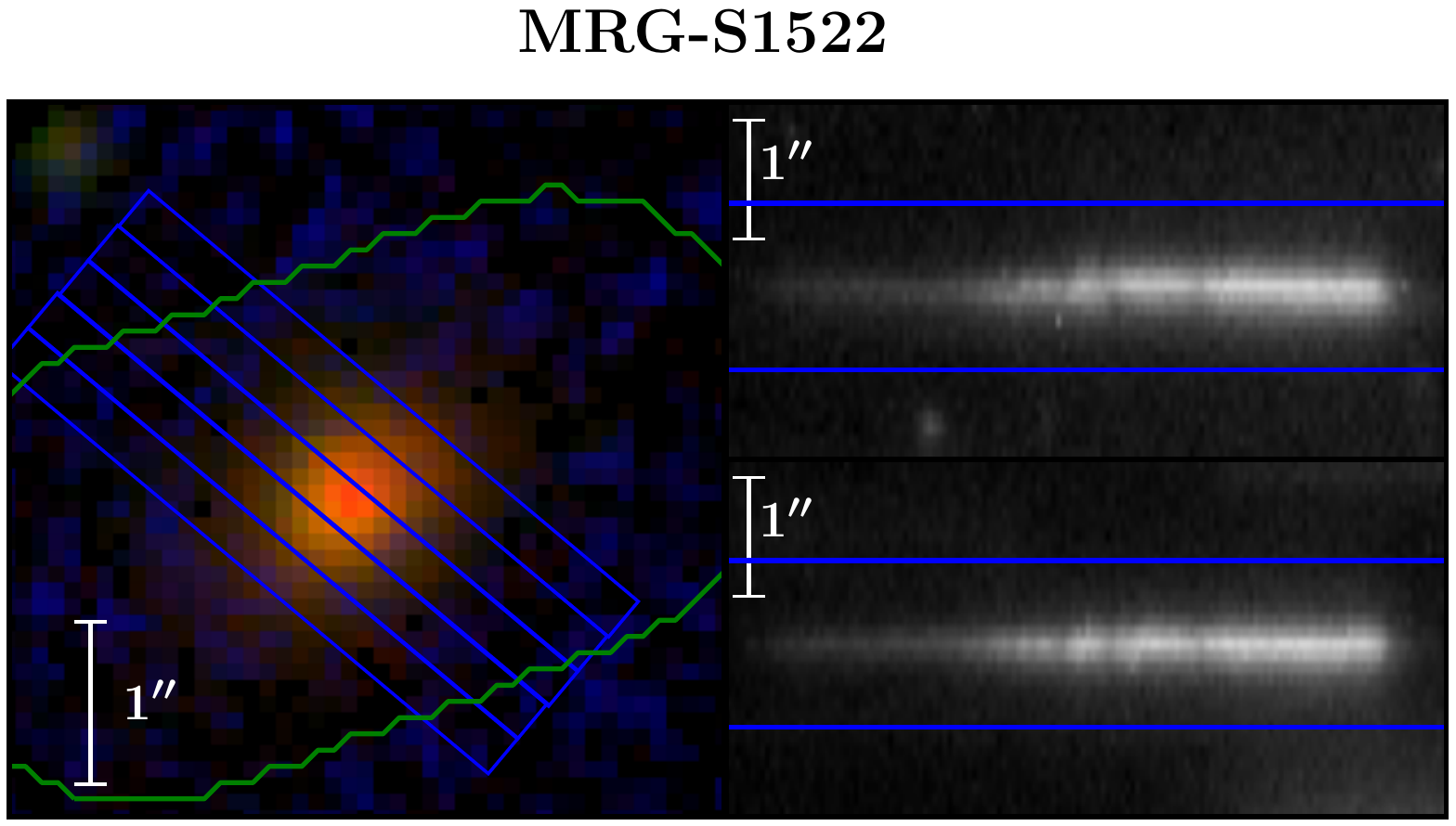}
&
\includegraphics[width=0.32\textwidth]{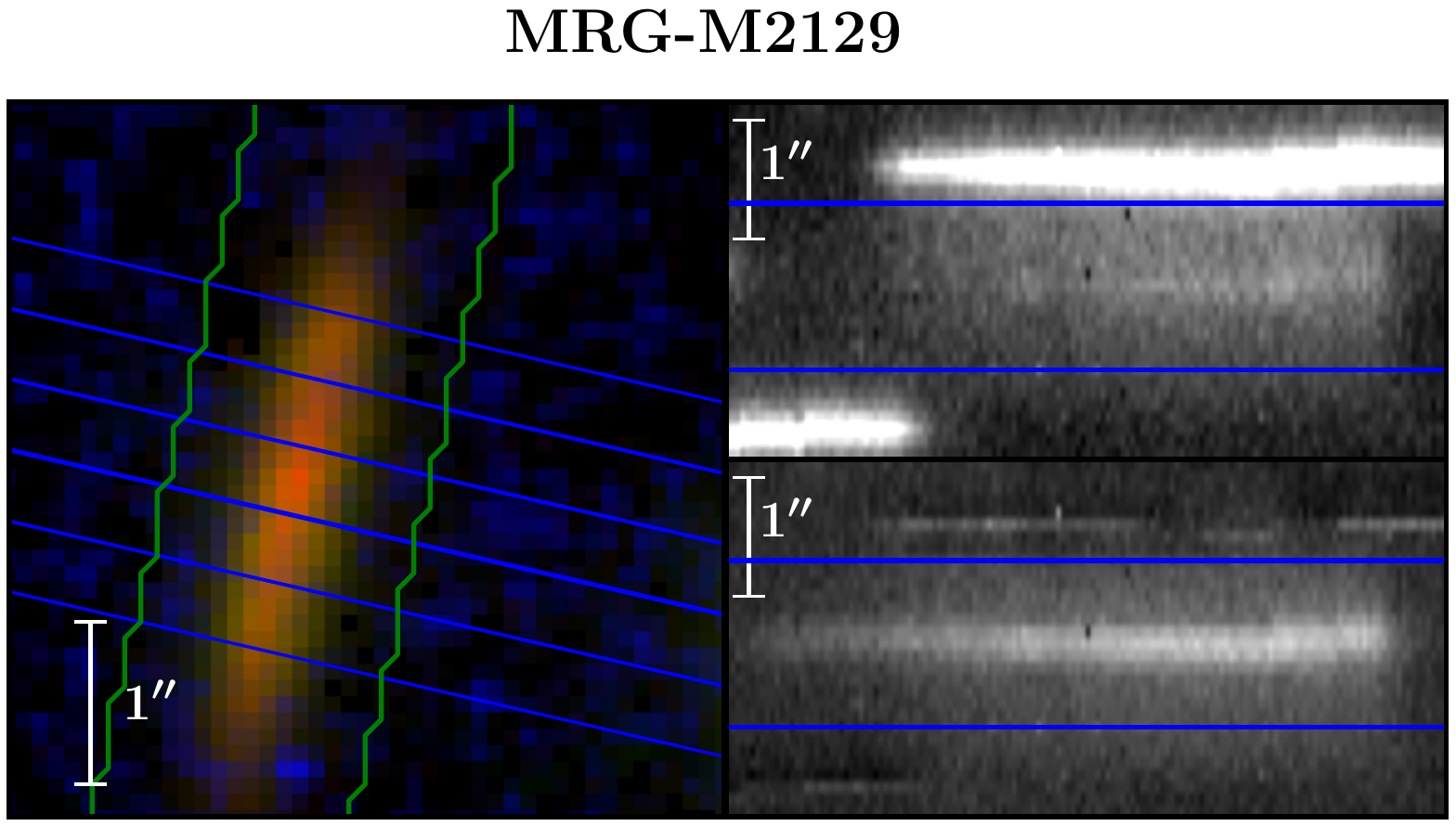}
\\
{} & \includegraphics[width=0.32\textwidth]{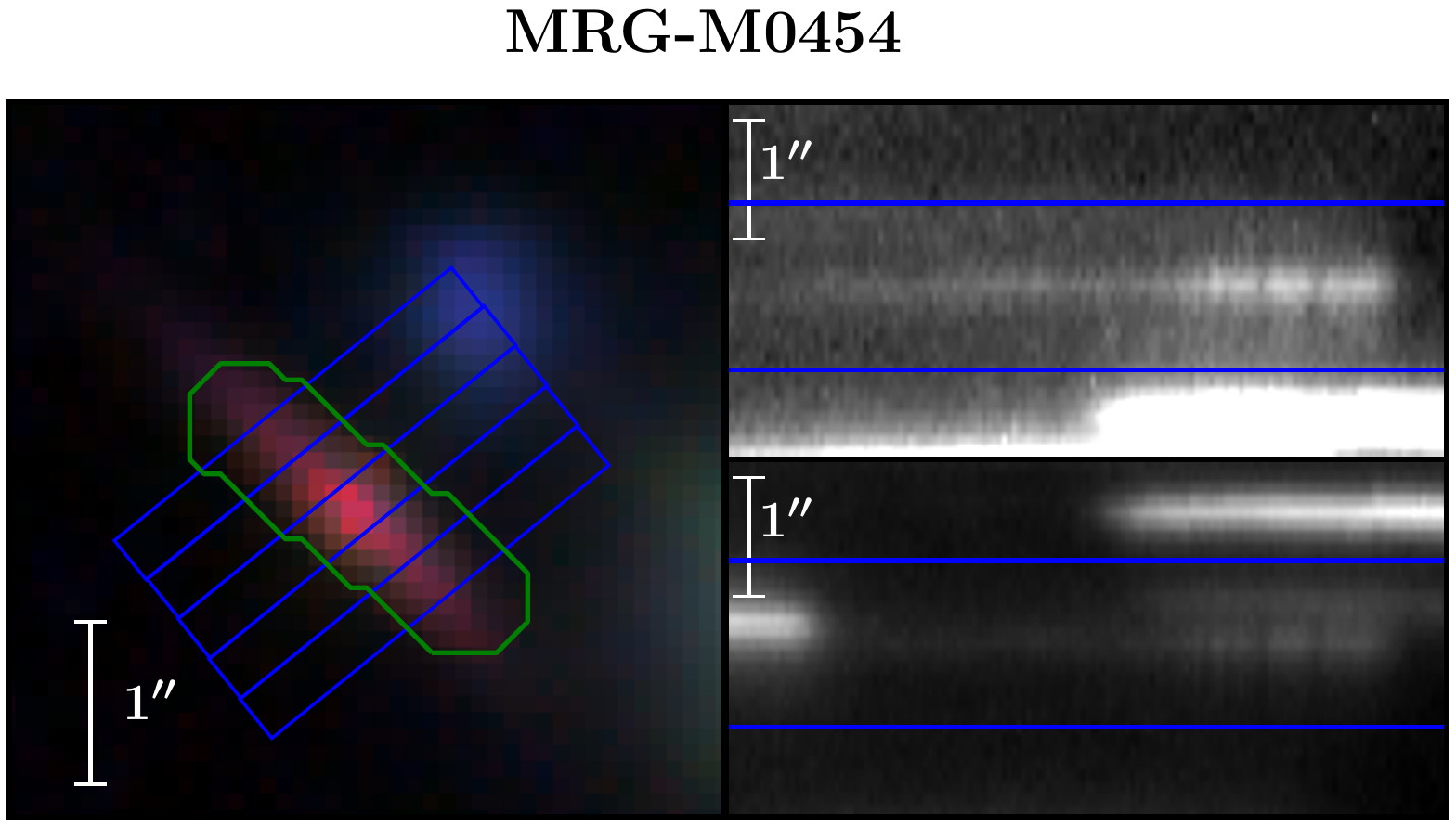}
& {}
\end{tabular}
\caption{Reduced \emph{HST} observations of the REQUIEM-2D sample. The color images (left) are constructed using $\mathrm{H_{F160W}}$ as red, either $\mathrm{Y_{F110W}}$ or $\mathrm{J_{F125W}}$ as green, and $\mathrm{U_{F390W}}$, or $\mathrm{I_{F814W}}$ or $\mathrm{V_{F555W}}$ as blue. The green contours show one of the flexible banana apertures that we use to measure photometric fluxes. The blue lines in each sub-panel on the left define the five central bins used in the spatially-resolved analysis. The two right sub-panels show data for the two grism dispersion angles used to observe each target in order to minimize contamination. The wavelength is increased from left to right. The blue lines in the right sub-panels represent the equivalent extent of the 5 central bins. \label{fig:color-im}}
\end{figure*}

\section{Data Reduction and Photometric Measurements} \label{sec:data_red}
\subsection{Data Reduction}
\begin{table*}[!t]
    \centering
    \begin{tabular}{|c|c|c|c|c|c|c|c|}
    \hline
filter/target &MRG-M1341 &MRG-M0138 &MRG-M0150 &MRG-P0918 &MRG-S1522 &MRG-M2129 &MRG-M0454 \\ \hline\hline 
F336W &$\times$ &$\times$ &$\times$ &$\times$ &$\times$ &12$\pm$7$\times10^{ -19 }$ &$\times$ \\ \hline 
F390W &$\times$ &$\times$ &$\times$ &$\times$ &7$\pm$4$\times10^{ -19 }$ &7$\pm$5$\times10^{ -19 }$ &$\times$ \\ \hline 
F555W &$\times$ &4$\pm$2$\times10^{ -18 }$ &$\times$ &11$\pm$5$\times10^{ -19 }$ &$\times$ &5$\pm$2$\times10^{ -19 }$ &54$\pm$3$\times10^{ -20 }$ \\ \hline 
F606W &33$\pm$4$\times10^{ -19 }$ &$\times$ &6$\pm$3$\times10^{ -19 }$ &$\times$ &10$\pm$2$\times10^{ -19 }$ &5$\pm$2$\times10^{ -19 }$ &$\times$ \\ \hline 
F775W &$\times$ &$\times$ &$\times$ &$\times$ &$\times$ &9$\pm$2$\times10^{ -19 }$ &113$\pm$5$\times10^{ -20 }$ \\ \hline 
F814W &82$\pm$3$\times10^{ -19 }$ &$\times$ &12$\pm$2$\times10^{ -19 }$ &18$\pm$3$\times10^{ -19 }$ &12$\pm$3$\times10^{ -19 }$ &8$\pm$1$\times10^{ -19 }$ &108$\pm$2$\times10^{ -20 }$ \\ \hline 
F850LP &$\times$ &$\times$ &$\times$ &$\times$ &$\times$ &11$\pm$2$\times10^{ -19 }$ &100$\pm$6$\times10^{ -20 }$ \\ \hline 
F105W &215$\pm$2$\times10^{ -19 }$ &244$\pm$6$\times10^{ -19 }$ &$\times$ &26$\pm$2$\times10^{ -19 }$ &17$\pm$1$\times10^{ -19 }$ &157$\pm$9$\times10^{ -20 }$ &101$\pm$3$\times10^{ -20 }$ \\ \hline 
F110W &238$\pm$3$\times10^{ -19 }$ &36$\pm$1$\times10^{ -18 }$ &201$\pm$10$\times10^{ -20 }$ &38$\pm$1$\times10^{ -19 }$ &245$\pm$10$\times10^{ -20 }$ &258$\pm$8$\times10^{ -20 }$ &105$\pm$2$\times10^{ -20 }$ \\ \hline 
F125W &271$\pm$2$\times10^{ -19 }$ &43$\pm$1$\times10^{ -18 }$ &234$\pm$9$\times10^{ -20 }$ &475$\pm$5$\times10^{ -20 }$ &292$\pm$8$\times10^{ -20 }$ &328$\pm$7$\times10^{ -20 }$ &114$\pm$2$\times10^{ -20 }$ \\ \hline 
F140W &271$\pm$1$\times10^{ -19 }$ &497$\pm$7$\times10^{ -19 }$ &403$\pm$7$\times10^{ -20 }$ &64$\pm$1$\times10^{ -19 }$ &418$\pm$7$\times10^{ -20 }$ &416$\pm$6$\times10^{ -20 }$ &172$\pm$2$\times10^{ -20 }$ \\ \hline 
F160W &266$\pm$1$\times10^{ -19 }$ &525$\pm$6$\times10^{ -19 }$ &548$\pm$7$\times10^{ -20 }$ &72$\pm$2$\times10^{ -19 }$ &525$\pm$7$\times10^{ -20 }$ &425$\pm$5$\times10^{ -20 }$ &238$\pm$1$\times10^{ -20 }$ \\ \hline 
IRAC1 &106$\pm$5$\times10^{ -19 }$ &34$\pm$1$\times10^{ -18 }$ &32$\pm$2$\times10^{ -19 }$ &27$\pm$4$\times10^{ -19 }$ &22$\pm$1$\times10^{ -19 }$ &209$\pm$6$\times10^{ -20 }$ &100$\pm$4$\times10^{ -20 }$ \\ \hline 
IRAC2 &68$\pm$2$\times10^{ -19 }$ &240$\pm$6$\times10^{ -19 }$ &23$\pm$2$\times10^{ -19 }$ &$\times$ &157$\pm$8$\times10^{ -20 }$ &147$\pm$4$\times10^{ -20 }$ &69$\pm$3$\times10^{ -20 }$ \\ \hline

   \end{tabular}
    \caption{Photometric measurements of the total flux density ($F_\lambda$ [erg/s/$\mathrm{cm}^2$/$\AA$]) for each REQUIEM-2D target used in our study, including \emph{HST} WFC3, ACS and \emph{Spitzer} IRAC filters (see Section \ref{sec:data_red}). Whenever a filter has not been observed for a target, we denote it by $\times$. MRG-M2129 also has overlapping F275W and F225W exposures not listed here; it is undetected in these filters with a  1$\sigma$ upper limit of $1.1\times 10^{-18}$ and $1.7\times 10^{-18}$, respectively$^a$. The measured 1.3 mm ALMA $F_\lambda$ flux densities are $48\pm5\times 10^{-24}$ (MRG-M0138) and $173\pm3\times10^{-24}$ (MRG-M2129) in erg/s/$\mathrm{cm}^2$/$\AA$ \citep{whitaker2021}. \newline
    \scriptsize ${}^a$ The \emph{HST} observing programs are GO-14622, PI: Whitaker; GO-15663, PI: Akhshik; GO-15466, PI: Ebeling; GO-13459, PI: Treu; PI: Ebeling; SNAP-14098, PI: Ebeling; SNAP-12884, PI: Ebeling; SNAP-15132, PI: Ebeling; SNAP-11103, PI: Ebeling; GO-12099, PI: Riess; GO-12100, PI: Postman; GO-9722, PI: Ebeling; GO-9292, PI: Ford; GO-10493, PI: Gal-Yam; GO-14496, PI: Newman; GO-14205, PI: Newman; GO-13003, PI: Gladders; GO-11591, PI: Kneib; GO-9836, PI: Ellis.}
    \label{tab:fluxes}
\end{table*}

The REQUIEM-2D galaxy survey targets 8 strongly lensed quiescent galaxies with redshifts of $1.6<z<2.9$ and demagnified stellar masses of $10.6<\log M_*/M_\odot<12$ \citep[see Figures \ref{fig:sfms} and \ref{fig:color-im} and also,][HST-GO-15663; PI: M. Akhshik]{akhshik2020}. The pilot target of the survey, MRG-S0851 (HST-GO-14622; PI: K. E. Whitaker), has been studied in \citet{akhshik2020, akhshik2021} and \citet{caliendo2021}. Therein, we built and extensively tested our methodology on mock observations as well as the pilot target. We closely follow this methodology to analyze the remaining 7 galaxies in this paper:

\begin{itemize}
     \item MRG-M1341: a highly magnified $\mu\sim 30$ galaxy at $z=1.59$ ($\mu$ is the linear gravitational magnification) \citep{ebeling2018, whitaker2019} (15 orbits of WFC3/G141),
    \item MRG-M0138: a very bright, massive galaxy at $z=1.95$ with $\mathrm{H_{F160W}}=17.3$ \citep{newman2018} (6 orbits of WFC3/G141),
    \item MRG-M2129 and MRG-M0150: massive quenched galaxies at $z=2.1$ and $z=2.6$ that are rotationally-supported and dispersion-dominated, respectively \citep{newman2015, toft2017} (5 orbits of WFC3/G141 each),
    \item MRG-P0918 and MRG-S1522: $\sim 1$ Gyr old quiescent galaxies at $z=2.36$ and $z=2.45$ \citep{newman2018} (7 orbits of WFC3/G141 each),
    \item MRG-M0454: the most compact $r_{\mathrm{eff}}\sim 0.3$kpc and the highest redshift $z=2.9$ galaxy in the survey \citep{man2021} (12 orbits of WFC3/G141).
  \end{itemize}
  
We use the Mikulski Archive for Space Telescopes (MAST) to access all publicly available \emph{HST} Wide Field Camera 3 (WFC3) and Advanced Camera for Surveys (ACS) data (see Table \ref{tab:fluxes}) along with the REQUIEM-2D Cycle 26 \emph{HST} program (HST-GO-15663). All \emph{HST} and \emph{Spitzer} data is reduced using the ``Grism redshift \& line analysis software for space-based slitless spectroscopy'', or \texttt{Grizli} \citep{brammer2016}. \texttt{Grizli} performs astrometric calibrations of the WFC3-IR and WFC3-UVIS images using the Pan-STARSS \citep{flewelling2016} and the Gaia-DR2 catalog \citep{gaia2018a,gaia2018b}. The final drizzled mosaics are constructed using \texttt{AstroDrizzle}\citep{avila2012} with a \texttt{pixfrac} of 0.33 and a pixel size of $0\farcs1$.

\subsection{Constraining Photometric Fluxes and Morphologies \label{subsec:photometry}}

After reducing \emph{HST} and \emph{Spitzer} data, we measure photometric fluxes that are major components of our joint spectrophotometric fitting. We first construct \emph{HST} point spread functions (PSFs) using \texttt{Grizli}. In this step, \texttt{Grizli} registers predefined empirical PSFs, retrieved from MAST \citep{anderson2016}, of each filter for the existing mosaics using the spatial resolution and orientation angles of the contributing exposures. We also construct \emph{Spitzer} PSFs in a similar way to the \emph{HST} PSFs. To build our \emph{Spitzer} PSFs, we first utilize the rich data set in the GOODS-S field and construct the `base’ empirical PSFs in each filter in a 0\farcs1 pixel scale. Then, we mimick the visits of the telescope over the REQUIEM fields, building our position-dependent PSFs on a $5^{\prime\prime} \times 5^{\prime\prime}$ grid, in a box with (a minimum of) $30^{\prime\prime}$ on a side around the target.

Using drizzled mosaics and weight maps, we measure the \emph{Spitzer} and \emph{HST} photometric fluxes and uncertainties using a methodology combining earlier works by \citet{akhshik2020}, \citet{wuyts2010}, and \citet{skelton2014}. For all targets, we construct a \texttt{Galfit} \citep{peng2002} model in the $\mathrm{H_{F160W}}$ band for a $20^{\prime\prime}\times20^{\prime\prime}$ box around each target, modeling all galaxies that can potentially contaminate our target (for MRG-M0138 the box is $37\farcs5\times 19^{\prime\prime}$ to accommodate its extended morphology). In each box, we have several light profiles, modeled using different \texttt{Galfit} components, and they are used to clean the contamination in all other \emph{HST} bands after matching the PSF, allowing them to have different overall flux normalizations. We finally use these models to remove the light profiles of all surrounding objects from the images.

Conventional circular-aperture-photometric methods fail to provide a reliable estimate for most targets in the REQUIEM-2D survey due to their extended and somewhat irregular morphologies. We therefore use a variation of ``object apertures'' following \citet{wuyts2010}: the axis of symmetry is determined by fitting a polynomial function to isophotes of each galaxy, requiring that the distances of opposing points from the axis of symmetry be minimum. We inspect each fit to make sure that the axis of symmetry is satisfactory. Apertures with increasing radii are centered on this axis defining banana-shaped apertures. These apertures are then used to measure the flux densities (see Figure \ref{fig:color-im}). 

It is a standard practice to measure the flux density using a smaller $0\farcs7$ diameter aperture and implement a correction factor to total \citep[e.g.,][]{skelton2014}, as using larger apertures does not optimize the signal to noise ratio. Thus, we pick an aperture with a radius of $0\farcs35\sqrt{\overline{\mu}}$, where $\overline{\mu}$ is the average linear gravitational magnification of the different images of the galaxy should it be multiply-imaged\footnote{Using smaller apertures are not suitable when the galaxy has color gradients.  The radius here is only measuring distances perpendicular to the axis of symmetry and the full size of the banana aperture is indeed larger (see Figure \ref{fig:color-im}). We also use the average gravitational magnification to avoid making the aperture large perpendicular to the axis of symmetry in case of extreme magnifications such as MRG-M1341.}. The aperture correction is then defined to be the ratio of the flux measured using the same aperture with a radius of $0\farcs35\sqrt{\overline{\mu}}$ on the noise-free \texttt{Galfit} model to the total flux.

To estimate the uncertainties of the \emph{HST} aperture photometry, we first define a set of empty apertures with increasing radii for the different filters of each target, measuring their fluxes. We then fit a power-law function to the width of the flux distribution and use this scaling relation to estimate the noise following \citet{whitaker2011}. For the \emph{Spitzer} IRAC channels, we use the square root of the sum of the inverse weight-map pixels as the uncertainty.

The morphology and the lensing models of MRG-M0138, MRG-M2129, MRG-M0150, MRG-M0454 and MRG-M1341 are studied thoroughly in the literature \citep{zitrin2014, umetsu2014, toft2017, newman2018,  sharon2020,  zalesky2020, jauzac2021, man2021, rodney2021}. MRG-S1522 has a published lensing model, too \citep{sharon2020}. The lensing model for MRG-P0918 is presented in Appendix \ref{sec:lensing}.

\section{Spatially Resolved Stellar Populations}

\subsection{Analysis\label{sec:resolved-analysis}}

We analyze spectrophotometric data using the \texttt{requiem2d} code\footnote{publicly available at https://github.com/makhshik/requiem2d}, referring the reader to \citet{akhshik2020} for more details. The fully Bayesian model is able to constrain the spatially resolved ages and star-formation histories (SFHs) of massive distant galaxies using spectrophotometric observations that assume a non-parametric SFH. We choose the grism dispersion angle such that it is perpendicular to the axis of symmetry while also minimizing the contamination from nearby objects.  We define 7 spatially resolved bins per galaxy using the $\mathrm{H_{F160W}}$ image for our analyses following \citet{akhshik2020} (see Figure \ref{fig:color-im}). Each one of the five central bins are defined to be 3-4 pixels wide in the image plane ($0\farcs3$-$0\farcs4$) noting that this resolution is greater than the FWHM of the PSF of $0\farcs18$. The two outer bins include all of the remaining pixels within the segmentation map on each side.

One important difference in the analyses here and that of \citet{akhshik2020} is the number of time bins in the non-parametric SFH model, i.e. the lookback time resolution. \citet{akhshik2020} assume the original FSPS \citep[Flexible Stellar Population Synthesis;][]{conroy2009, conroy2010} time resolution ($\sim$ 70 time bins), whereas we find that this approach introduces some artifacts in the recovered SFHs that are produced by different built-in FSPS assumptions. One notable example is the age when \texttt{dust1} is added to the model: all stars younger than this time scale, with a default value of $10^7$ years, are attenuated by both \texttt{dust1} and the usual \texttt{dust2} \citep{conroy2010}. It can be shown that at around this time scale, some slight enhancements/depressions manifest themselves in the SFH \citep[e.g., Figure 1 in][around 0.01 Gyr]{akhshik2021}. We explicitly check this effect for MRG-P0918 by changing this time scale from 0.01 to 0.015 and 0.02 Gyr and refit the model. The artifacts move in a consistent manner appearing at the assumed \texttt{dust1} time scale. To avoid adding artifacts into the recovered SFH, we instead define 20 time bins for all samples in the model\footnote{These features are small in the sense that we do not see any difference in the inferred parameters of interest such as stellar mass, age and specific star-formation rate by changing the \texttt{dust1} time scale.}. We however note that the \texttt{requiem2d} code is fully capable of fitting the finest lookback time resolution ($\sim$ 70-90) produced by FSPS if required.

For those targets with ALMA 1.3 mm detections, MRG-M0138 and MRG-M2129, we carefully examine the distribution of the dust continuum emission. In both sources, the dust emission is unresolved in the $\sim$1-1\farcs5 ALMA data \citep{whitaker2021}. While this resolution is significantly coarser than our spatially resolved bins, the fact that the emission is pointlike implies that there is some information about the spatial distribution of the 1.3 mm flux in the existing data. We performed a number of tests to determine the level of dust emission in each resolved bin that would be consistent with the existing ALMA data, jointly fitting for the 1.3 mm flux density of a series of unresolved point sources with positions fixed to the centers of each of our resolved bins. This analysis suggests that for both MRG-M0138 and MRG-M2129, the 1.3 mm emission is associated with the three central bins, though the distribution of flux among these bins is essentially unconstrained. In other words, the existing ALMA data constrain the total dust emission in these sources and localize it to the central three bins in each source. Therefore, in our fitting procedure we sum the predicted 1.3 mm flux density of the central 3 bins and compare this to the observed 1.3 mm emission, while for the exterior bins we use only upper limits on the millimeter flux. We implement the ALMA semi-resolved flux in the Bayesian model using the integrated likelihood \citep[e.g., Section 4.3 of][]{stan2021}, i.e. for those bins with only an upper limit detection, the integral of the flux likelihood up that upper limit is included in the Bayesian fitting. For other targets that are undetected in ALMA 1.3 mm, we can in principle use the upper limit but as these targets are not particularly dusty (see Figure \ref{fig:gradients}), adding the upper limit would not be additionally constraining. 

\begin{figure*}[!t]
\gridline{\fig{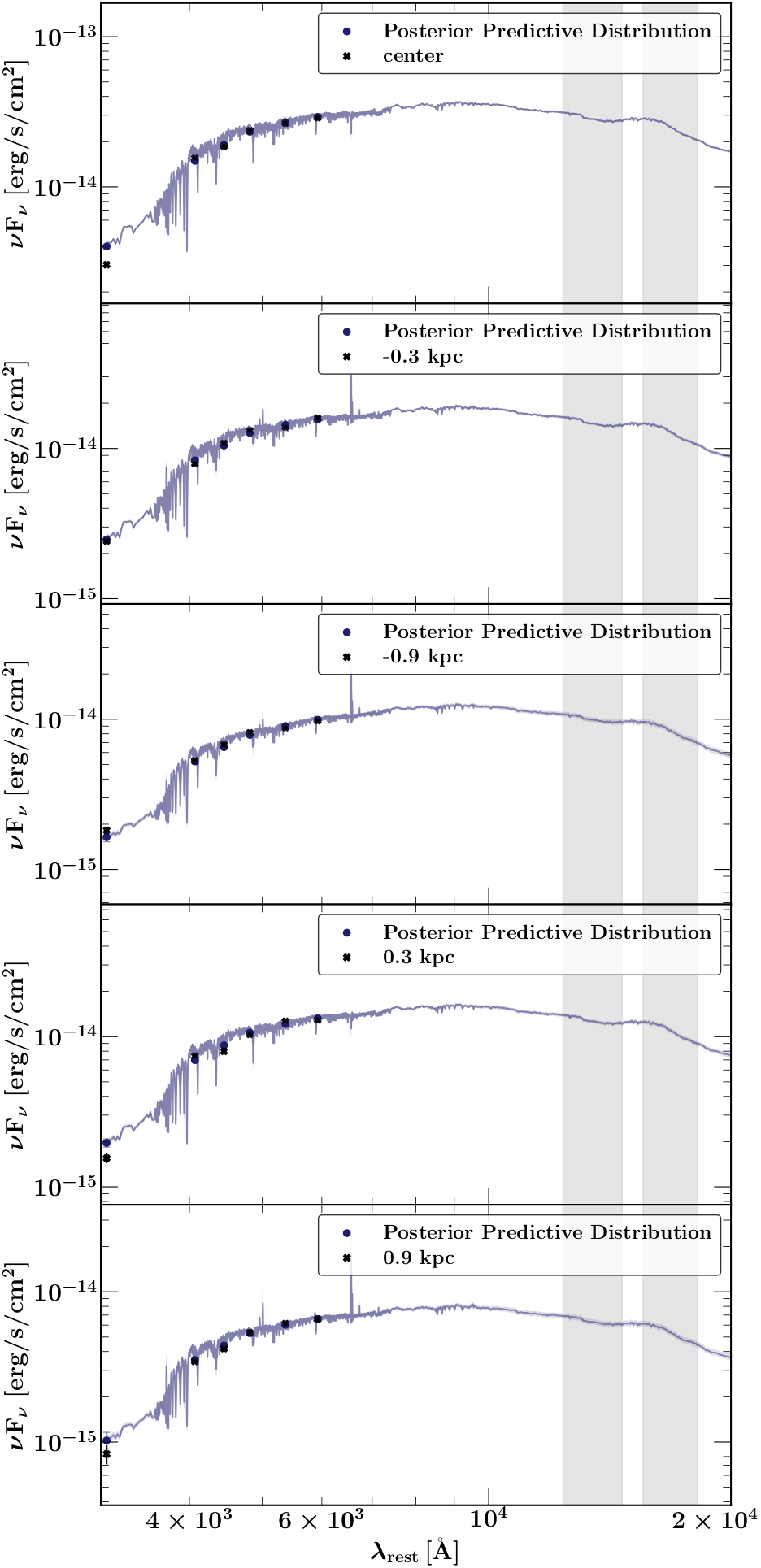}{0.45\textwidth}{}
\hspace{0in}\fig{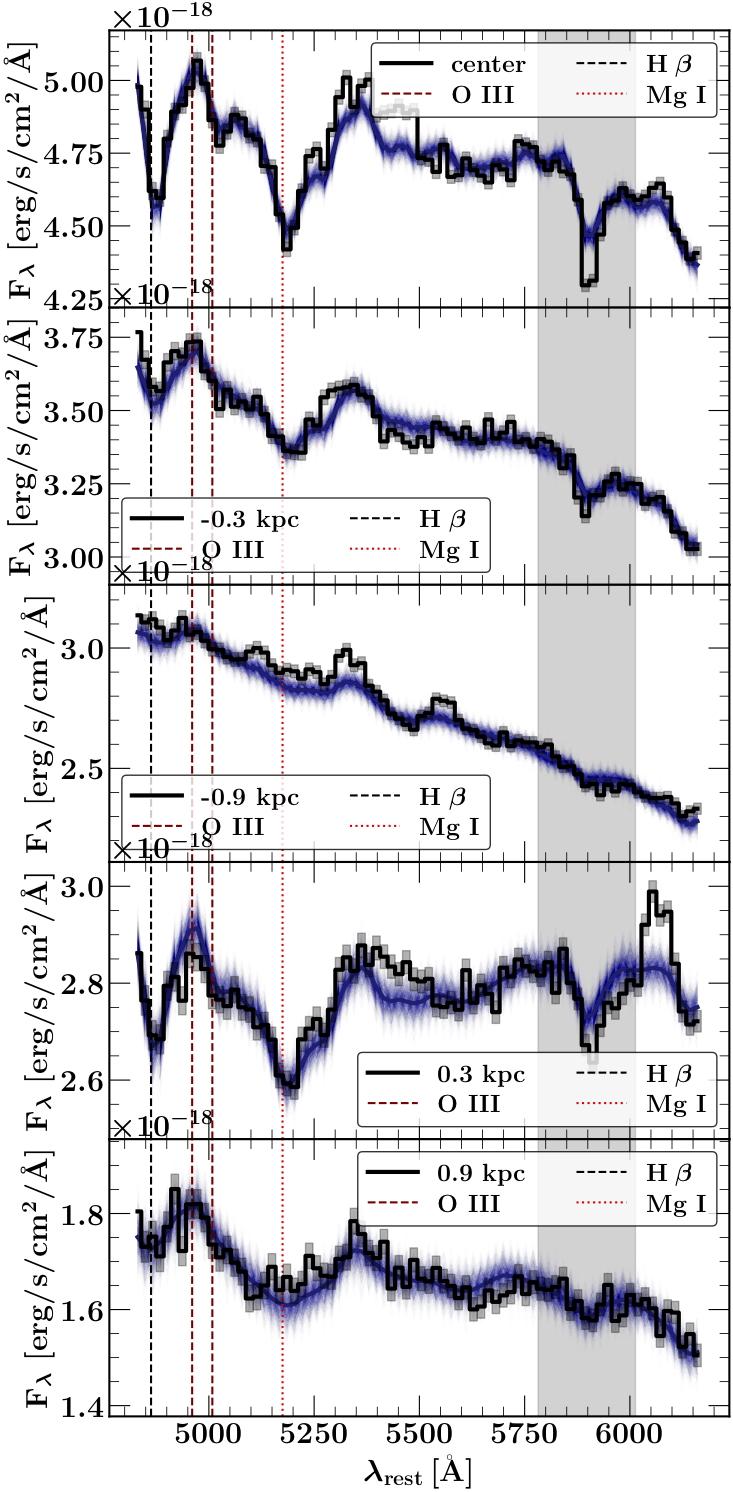}{0.465\textwidth}{}}
\vspace{-0.3in}
\caption{\textbf{Left:} The SED of MRG-M1341. The blue line shows 200 draws from the FSPS model. The shaded regions in the SED coincide with the unresolved \emph{Spitzer} IRAC channels 1 and 2 band passes. \textbf{Tight:} The resolved extracted 1D grism spectra. The shaded region (1.50-1.56 $\mu$m in the observed frame) in spectra is Na D line, which is masked in our fitting.  \label{fig:resolved-mrg-m1341}}
\end{figure*}

We use the \citet{noll2009} parameterization for dust, assuming a \citet{kriek2013} dust model \citep[e.g.,][]{akhshik2020}. The dust model has three main parameters: (1) \texttt{dust1} that describes the extra attenuation for stars younger than $10^7$ years, (2) \texttt{dust2}, describing the attenuation for stars older than $10^7$ years, and (3) \texttt{dust\_index}, that is the index of the power-law correction of the attenuation curve.

For each target we perform a separate stellar population fit using \citet[\texttt{Prospector-$\alpha$};][]{johnson2019,leja2017} to the semi-resolved photometric data, and subsequently use these posteriors in the joint spectrophotometric fit \citep{akhshik2020}. We divide the metallicity and \texttt{dust2} posterior (its 3$\sigma$ width) into 16 boxes, generating a series of FSPS templates for each box \citep{akhshik2020}. By refitting the weight of each box along with the SFHs using \texttt{requiem2d}, we practically use these \texttt{Prospector-$\alpha$} posteriors as weakly informative priors, and we do not expect this choice of priors affect our age estimates \citep[see Figure 15 in][for an example of starting from a different \texttt{Propsector-$\alpha$} posterior]{akhshik2020}. We adopt a slightly different approach to model MRG-M2129 and MRG-M0138. Because of the ALMA detections, we have non-trivial \texttt{Prospector-$\alpha$} posteriors for the dust emission parameters (\texttt{duste\_umin}, \texttt{duste\_qpah}, \texttt{duste\_gamma} in FSPS) along with \texttt{dust2} and metallicity. We therefore perform a principal component analysis for the full \texttt{Prospector-$\alpha$} posteriors and use two first principal components instead of \texttt{dust2} and metallicity to generate more representative FSPS templates. We also allow the dust attenuation curve to be fit separately, to ensure that the model is flexible enough to deviate from the \texttt{Prospector-$\alpha$} posterior to fit the spectrophotometric data. This step is implemented by defining half-normally distributed hyper-parameters, and using them as the standard deviations of the log-normally and normally distributed priors for \texttt{dust1}, \texttt{dust2} and \texttt{dust\_index}. Each one of these priors are centered at the corresponding mean values in each one of the 16 boxes and they are used to generate the attenuation curve. 

The No-U-Turn Sampler \citep[NUTS;][]{homan2014}, that is wrapped in \texttt{pymc3} \citep{salvatier2016}, is used to sample the full posterior. We generate two Monte Carlo chains, drawing 2000 samples from the posterior in each one (with the exception of older galaxies, MRG-M0138 and MRG-M1341, where we draw 3000 samples for each chain). We discard the first half of each chain as burn-in phase and compare the two chains to check for divergences using $\hat{R}$ \citep{gelman1992}. After ensuring convergence, the two chains are combined to construct a total of 2000 draws from the posteriors (3000 for MRG-M0138 and MRG-M1341).

\begin{figure*}[!t]
\centering
\vspace{-0.1in}
\begin{tabular}{cc}
\animategraphics[scale=0.06,controls,autoplay,loop]{0}{UVJ_gif_MRG-M1341_rbbg-}{0}{5}
&
\animategraphics[scale=0.06,controls,autoplay,loop]{0}{UVJ_gif_MRG-M0138_rbbg-}{0}{5}
\\
\animategraphics[scale=0.06,controls,autoplay,loop]{0}{UVJ_gif_MRG-M0150_rbbg-}{0}{5}
&
\animategraphics[scale=0.06,controls,autoplay,loop]{0}{UVJ_gif_MRG-P0918_rbbg-}{0}{5}
\\
\animategraphics[scale=0.06,controls,autoplay,loop]{0}{UVJ_gif_MRG-S1522_rbbg-}{0}{5}
&
\animategraphics[scale=0.06,controls,autoplay,loop]{0}{UVJ_gif_MRG-M2129_rbbg-}{0}{5}
\\
\animategraphics[scale=0.06,controls,autoplay,loop]{0}{UVJ_gif_MRG-M0454_rbbg-}{0}{5}
& {}
\end{tabular}
\caption{The SFH and the evolution of stellar mass, median age and U-V colors for different redshift snapshots. In the SFH panels, the central bin is shown by thick black line. The blue and red lines are the outskirt bins on each side, with the dashed line being the bin right next to the center bin. $r/r_c$ indicates the bin's distance from the center normalized to the distance that encloses half-light (see the text, Section \ref{sec:resolved-results} and Table \ref{tab:murc}). \label{animes}}
\end{figure*}

\subsection{Fits\label{sec:resolved-results}}

The \texttt{requiem2d} code provides the posterior of different parameters as well as the posterior predictive distributions. We show the spectral energy distributions (SEDs), the 1D collapsed grism spectra, and the posterior predictive distributions of each REQUIEM-2D target in Figure \ref{fig:resolved-mrg-m1341} here and Figures \ref{fig:resolved-mrg-m0138} to \ref{fig:resolved-mrg-m0454} in Appendix \ref{sec:plots}. 

The SED fit to the REQUIEM-2D sample is shown in the left panels of Figures \ref{fig:resolved-mrg-m1341} and \ref{fig:resolved-mrg-m0138}-\ref{fig:resolved-mrg-m0454}. We include the \emph{Spitzer} IRAC channels 1 and 2 as constraints by imposing likelihoods for the global fluxes that are normally distributed with the observed global values and their uncertainties as means and standard deviations and they are modeled by the summed fluxes of the individual bins. In other words, these data points are not spatially resolved like the \emph{HST} photometric bands. We therefore demonstrate their corresponding wavelength ranges with a grey shade in the SED plots.

The right panels of Figures \ref{fig:resolved-mrg-m1341} and \ref{fig:resolved-mrg-m0138}-\ref{fig:resolved-mrg-m0454} show the 1D collapsed spectra of the five central bins for each target, extracted following \citet{horne1986}. While we fit the two outer bins for all targets, the fit is not as reliable because of the fainter wings of the galaxy and/or contamination by nearby sources \citep[e.g.,][also see Figure \ref{fig:color-im}]{akhshik2020}. We therefore do not consider the outer bins when analyzing the radially-varying trends.

For each target, we demonstrate the important absorption lines that drive the age and SFH fit using vertical lines. One notable exception is Na D absorption line in MRG-M1341 (Figure \ref{fig:resolved-mrg-m1341}). We mask this line since it seems that the stellar absorption cannot fully explain its strength, especially at the center, potentially implying further absorption by the ISM. This is consistent with earlier observations of the Na D line in MRG-M0138 \citep{jafariyazani2020}\footnote{Na D line for MRG-M0138 is outside of the wavelength coverage of WFC3/G141.}. The masked region is shown with a shaded grey box in Figure \ref{fig:resolved-mrg-m1341}, right panel; the spectral model in this shaded box is not a fit but is rather an extrapolation. The tension between the data and the best fit model clearly demonstrates that the stellar absorption assumed for the rest of the model cannot fully explain the Na D strength.

The distances reported in Figures \ref{fig:resolved-mrg-m1341} and \ref{fig:resolved-mrg-m0138}-\ref{fig:resolved-mrg-m0454} are measured simply by finding the distance between the centers of corresponding adjacent bin in arcseconds. This distance is then translated into physical units using the standard cosmology, and finally it is divided by $\sqrt{\mu}$ to estimate the source plane distance. The linear magnification is fairly uniform for all of the targets. We also calculate the distance where half of the total flux is enclosed ($r_c$ in Figure \ref{animes}, see Table \ref{tab:murc} for the measured values and their uncertainties) when moving along the axis of symmetry (see Section \ref{subsec:photometry}) while adding all flux in the perpendicular direction. We prefer to use $r_c$ instead of the usual half-light radius, $r_\mathrm{eff}$ because it better represents our image-plane bins, that are in turn defined according to the grism observational constraints \citep{akhshik2020}. We nevertheless note that the values of $r_c$ are consistent within the uncertainties with the published half-light radii \citep{man2021, newman2018}, justifying this decision. We also use the term ``inner core'' to refer to $r<r_c$.

To better visualize the evolution of the REQUIEM-2D sample as a function of lookback time, we generate a series of animations that shows the evolution of the different parameters at different redshift/lookback time snapshots (Figure \ref{animes}). In each snapshot, we use the SFH to generate the SEDs and to derive different parameters. Two main assumptions are made here: (1) we assume that there is no significant evolution of dust, and (2) we assume that stars and dust in the galaxies follow non-evolving distributions. To first order, we know these assumptions do not hold, especially in the case of mergers or starbursts, or if the galaxy experiences a major episode of dust formation/destruction \citep[e.g.,][]{Li2019}. However, it is safe to assume that these animations are accurate representations of each galaxy in snapshots closer to the observed redshift; we include the earlier epochs for visual purposes.  The stellar population maps are superimposed over the color image of each target in the three bottom panels.

\section{Discussion \label{sec:discussion}}

\begin{figure*}[!th]
\vspace{-0.1in}
\centering
\begin{tabular}{cc}
\includegraphics[width=0.48\textwidth]{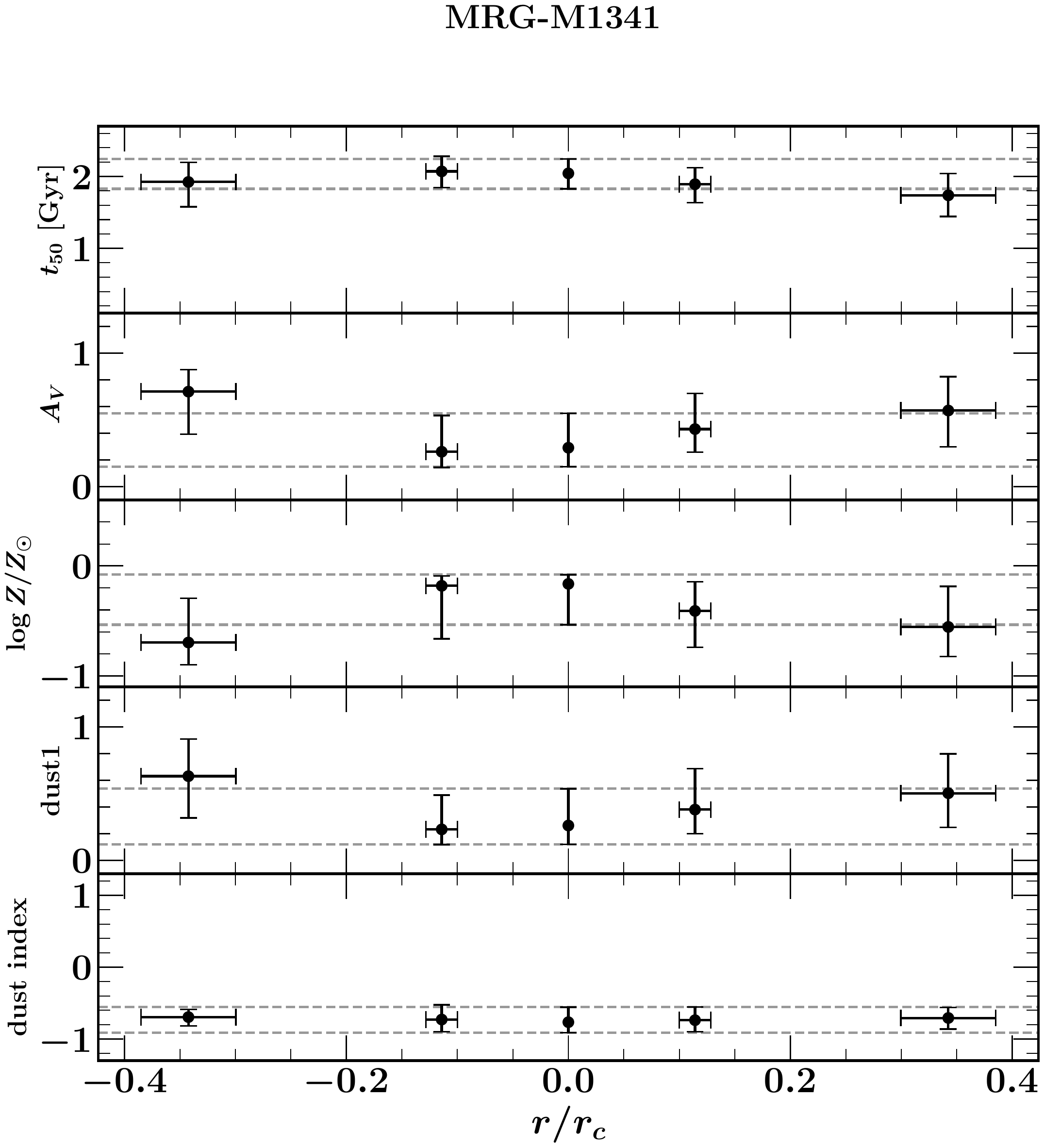}
&
\includegraphics[width=0.48\textwidth]{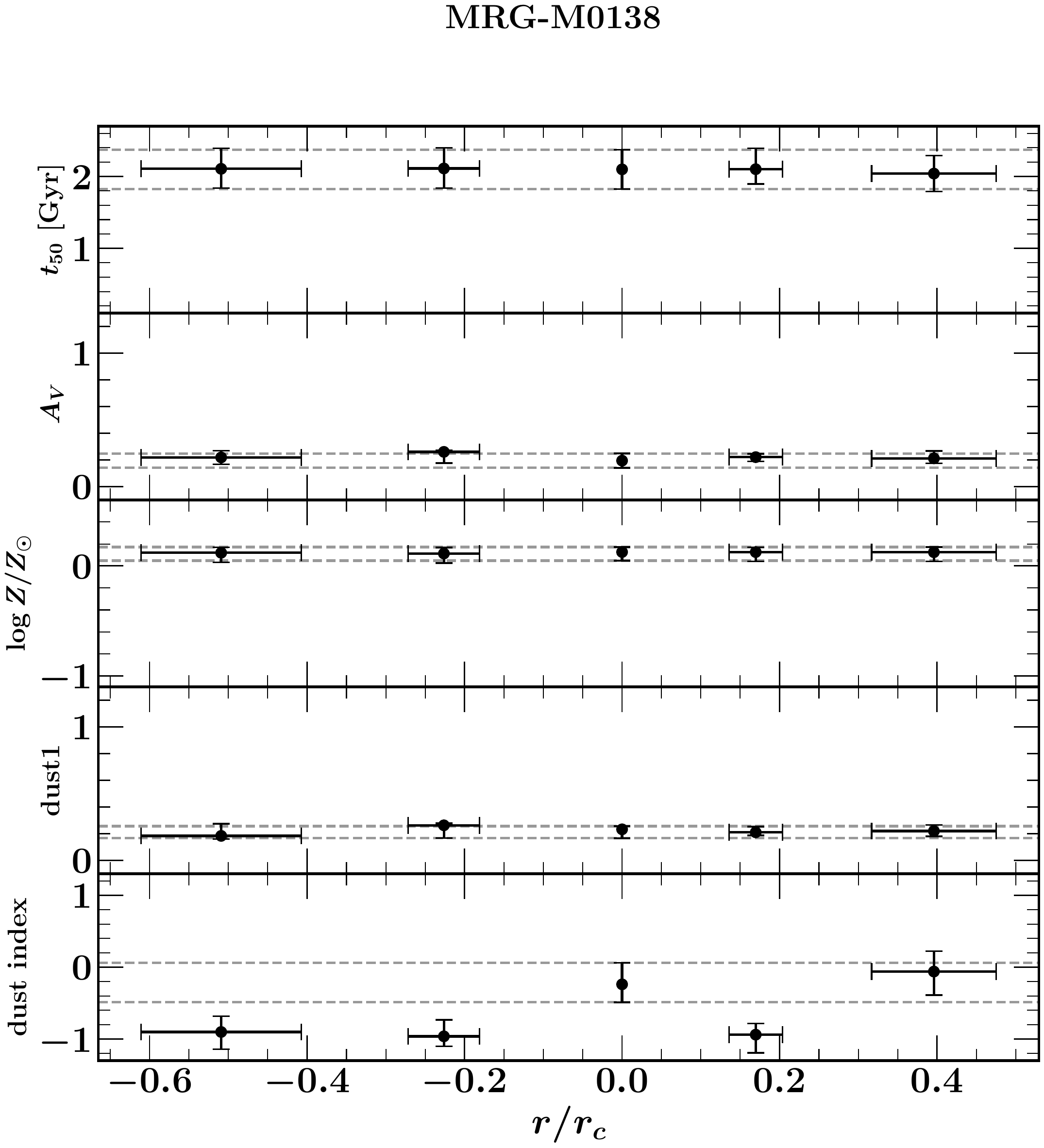}
\\
\includegraphics[width=0.48\textwidth]{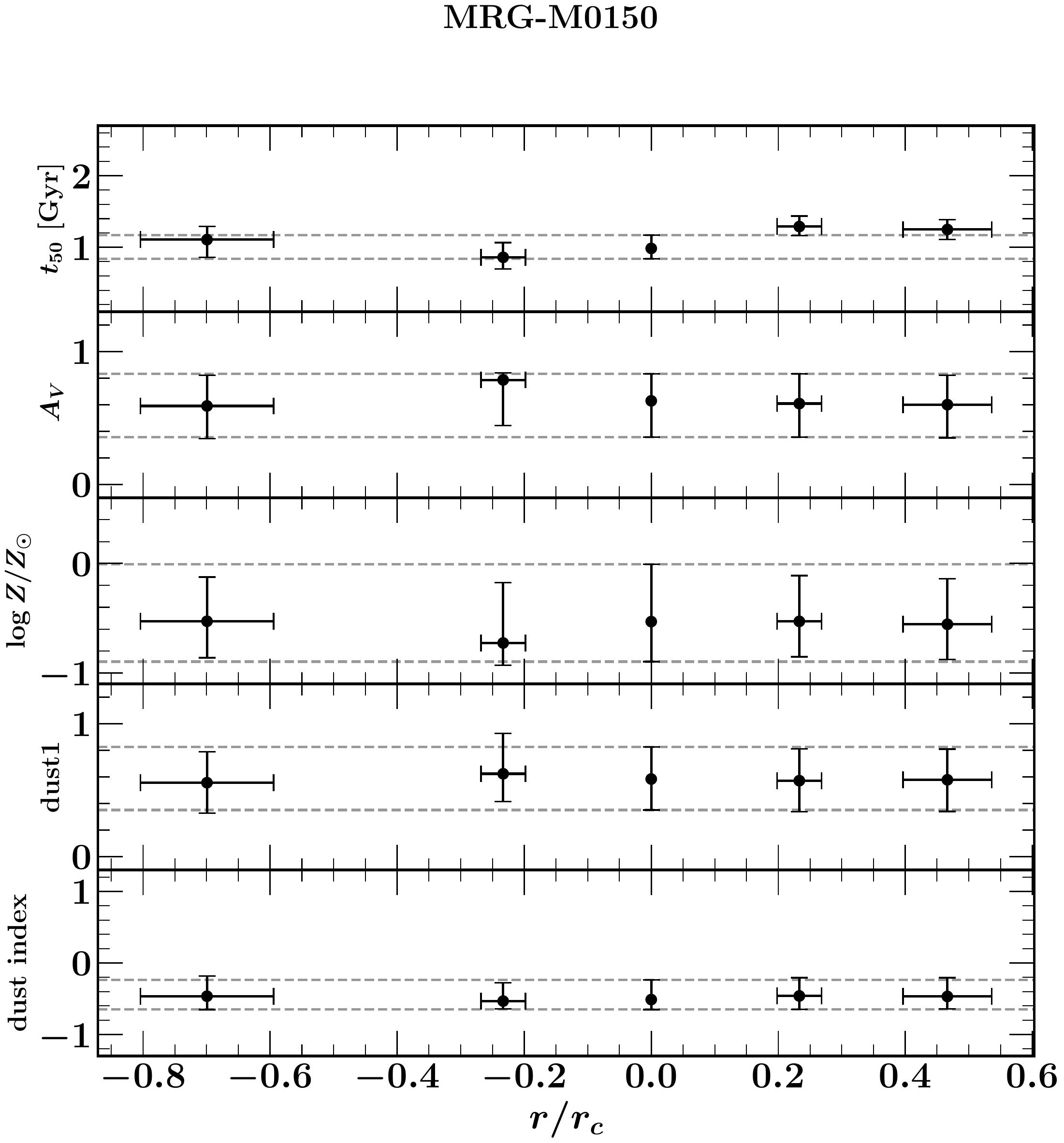}
&
\includegraphics[width=0.48\textwidth]{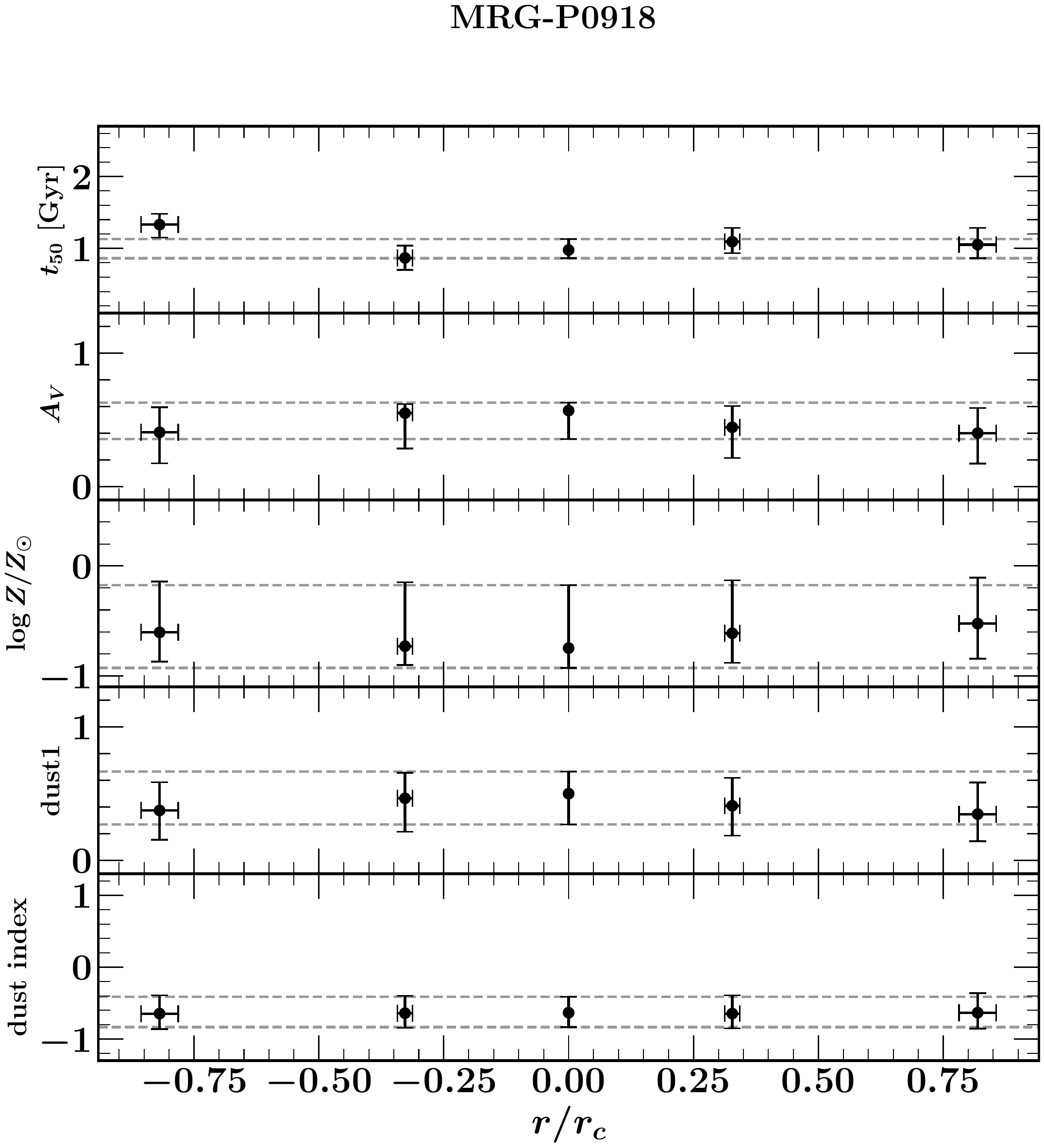}\\
\end{tabular}
\vspace{-0.1in}
\caption{Different stellar populations' parameters at the observing redshift. The dashed line shows the 1$\sigma$ width of the uncertainty for center bin, added to assist identifying potential gradients. Distances are the same as Figure \ref{animes}. MRG-S0815 results are from \citet{akhshik2020} and are added for completeness. \label{fig:gradients}}
\end{figure*}

\addtocounter{figure}{-1}

\begin{figure*}[!th]
\vspace{-0.1in}
\centering
\begin{tabular}{cc}
\includegraphics[width=0.48\textwidth]{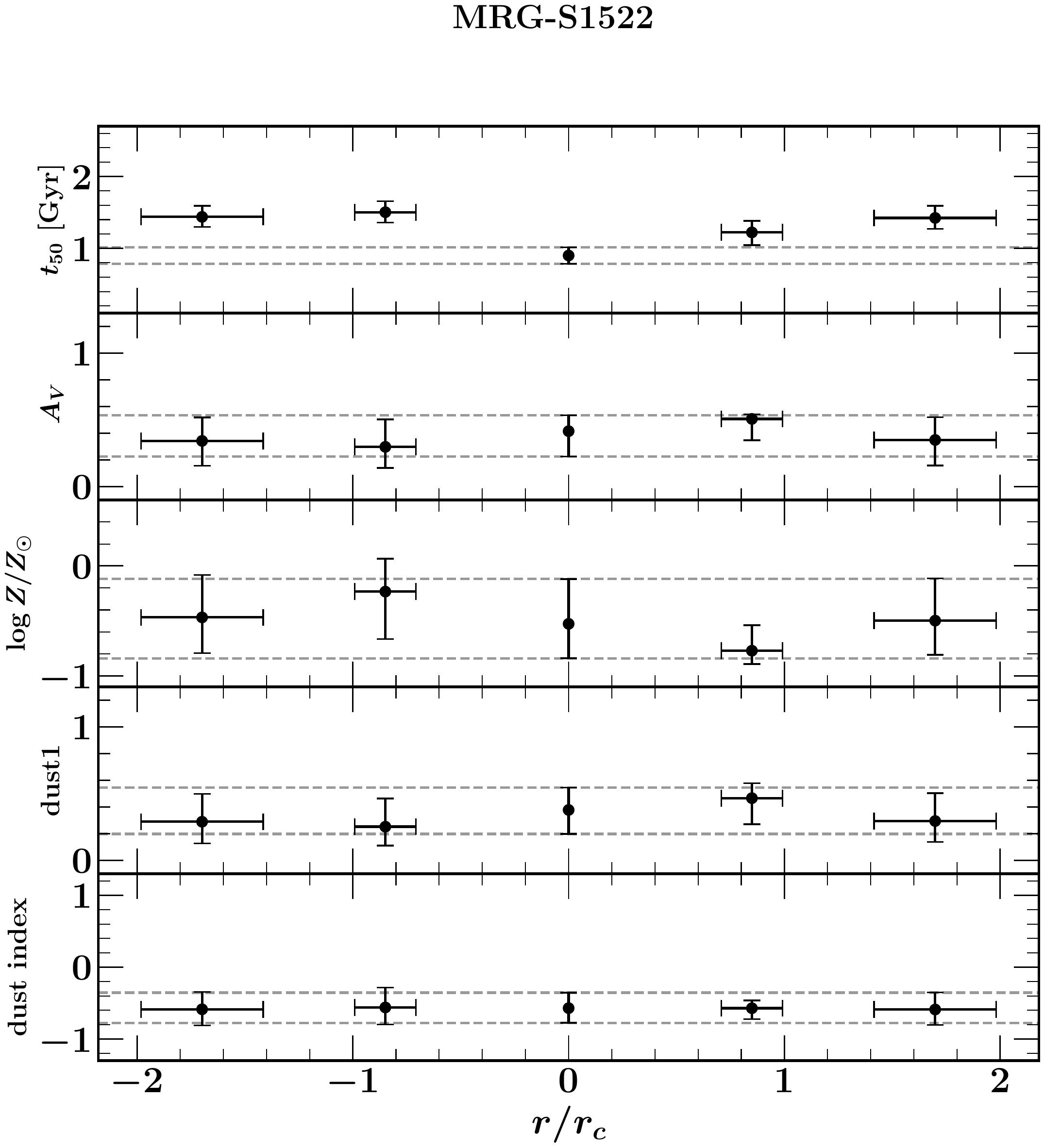}
&
\includegraphics[width=0.48\textwidth]{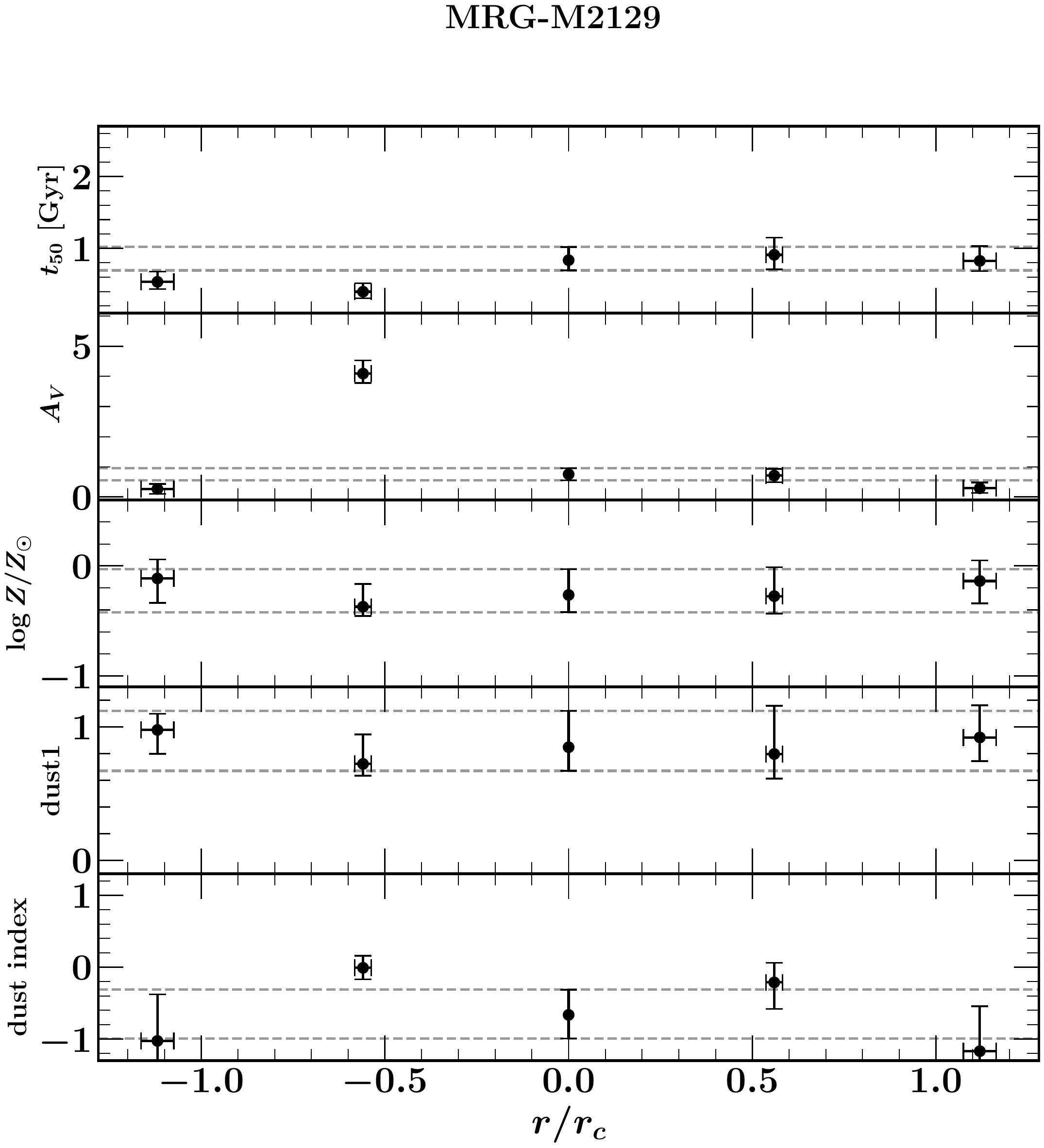}
\\
\includegraphics[width=0.48\textwidth]{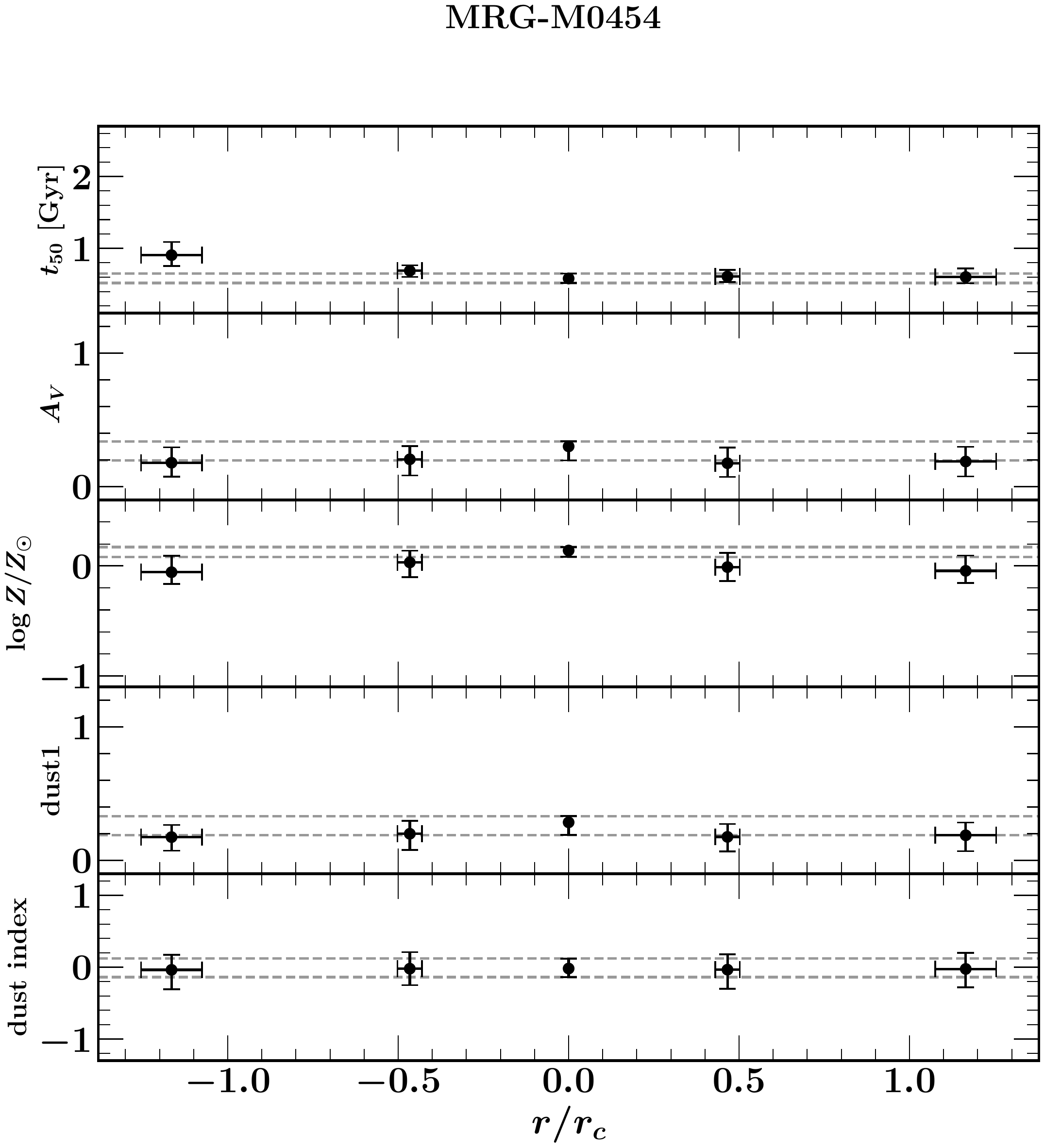}
& 
\includegraphics[width=0.48\textwidth]{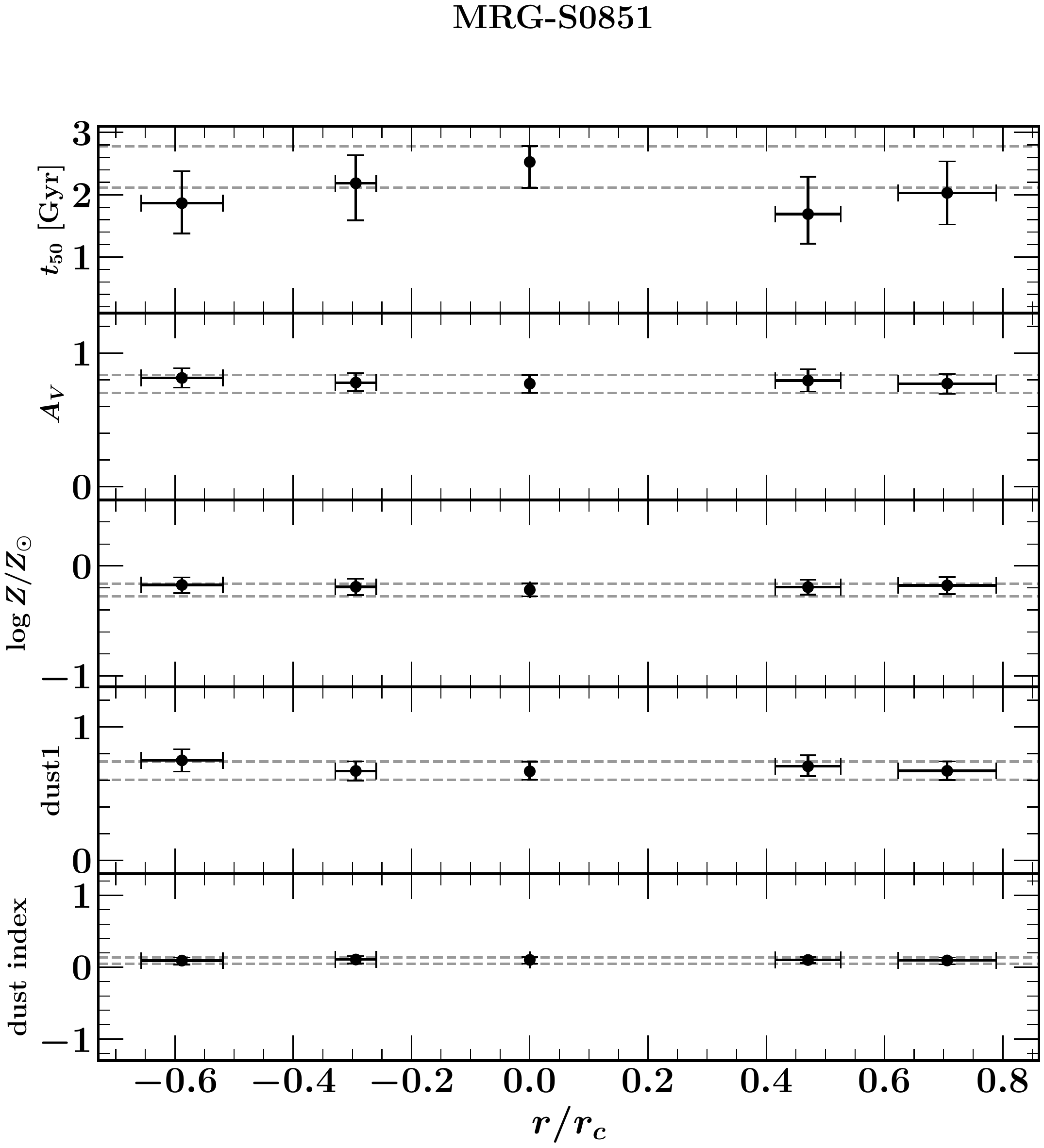}
\end{tabular}
\vspace{-0.1in}
\caption{continued}
\end{figure*}

\begin{figure*}[!t]
\centering
\includegraphics[width=\textwidth]{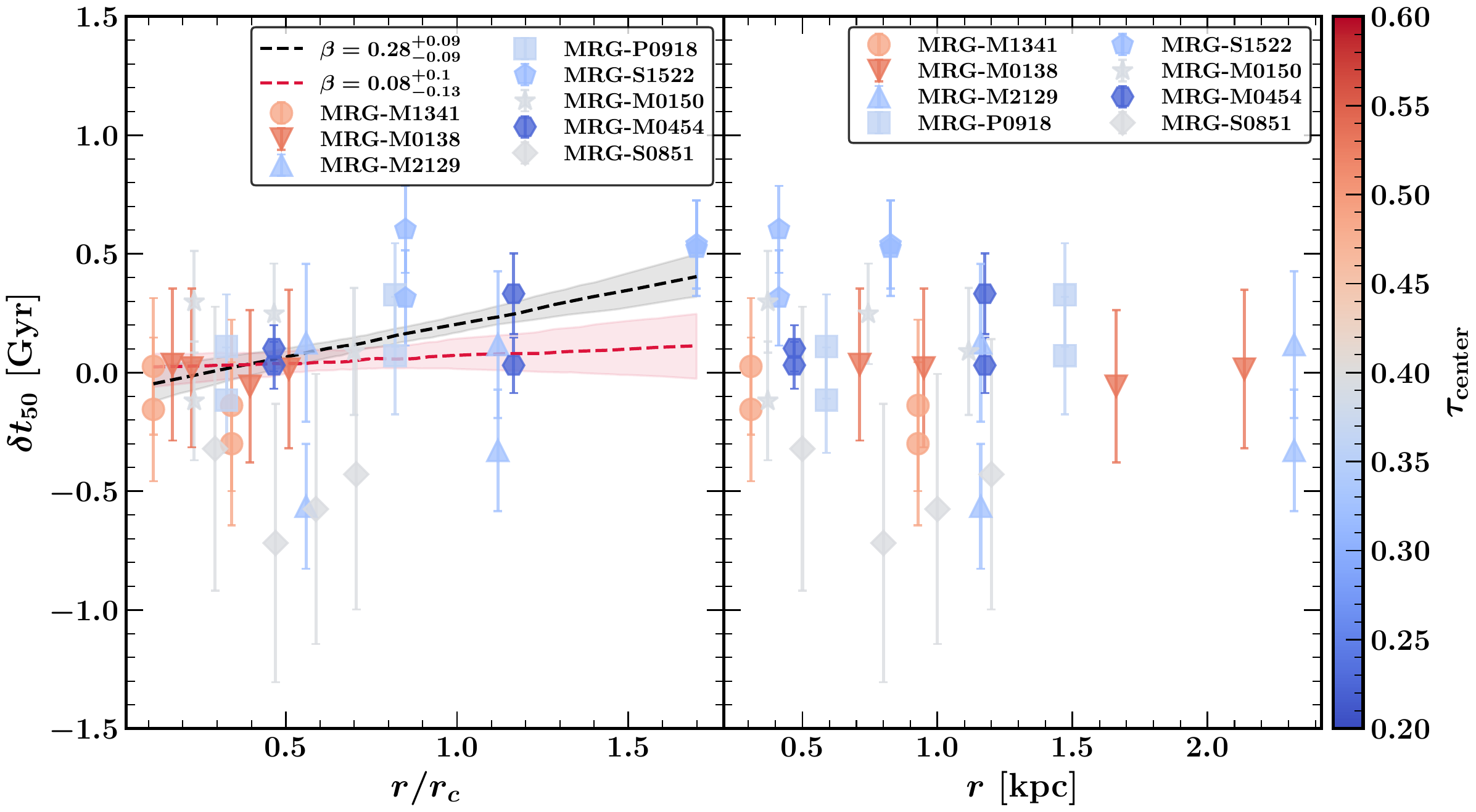}
\caption{Measured spatially resolved age differences for the REQUIEM-2D galaxies versus the distance and normalized distances to the half-light radii. $\delta t_{50}$ is the difference between the age of the corresponding bin and the center for each target in Gyr, i.e. $\delta t_{50}=t_{\mathrm{spatial\, bin}}-t_{\mathrm{center}}$ and the median age, $t_{50}$, is defined to be $0.5=\int_0^{t_{50}} dt \mathrm{SFR}(t)/\int_0^{t_0} dt \mathrm{SFR}(t)$, where $t_0$ is the observing lookback time. $r_c$ denotes the half light radius, and the black dashed line and shaded region are the best fit and the 1$\sigma$ width of the linear fit to all data points. The red dashed line is the same fit excluding MRG-S1522 data points, which is the only target showing a significant age difference beyond the half-light radius. $\beta$ denotes the slope of the linear fit and its 1$\sigma$ posterior uncertainty with a prior of $\beta\sim\mathcal{N}(0,100)$. Data points are color coded with the quenching time-scale of the central bin, defined following \citet{carnall2018}. \label{fig:age-grad}}
\end{figure*}

The goal of the REQUIEM-2D survey is to measure age and SFH gradients for the eight lensed quiescent galaxies through a spectrophotometric fit. We discuss our fitting algorithm in the preceding Sections, and here we present and discuss the implications of the measured gradients in the stellar populations.

We show the gradients of the different stellar population parameters in Figure \ref{animes} and \ref{fig:gradients}. While Figure \ref{animes} shows the evolution of parameters such as median age and sSFR for different redshift snapshots, Figure \ref{fig:gradients} demonstrates the values at the redshift of observation. We can identify three different categories of galaxies based on the detected age gradient patterns when noting the difference in the physical distances: (1) younger in the center, (2) flat, and (3) non-axisymmetric gradients. We choose mass-weighted age as the important parameter because it summarizes the average behavior of SFH. Also, it is suggested that age gradients can be connected to the formation pathway mechanisms of the galaxies \citep[e.g.,][]{wellons2015}.

MRG-S1522 is the only galaxy in the REQUIEM-2D sample which is clearly younger in the inner core ($r/r_c\lesssim 1)$. Its SFH in Figure \ref{animes} reveals that the center bin remains more star-forming relative to the outer bins between lookback times of $\sim$300 Myr and $\sim$1 Gyr, explaining the younger age at the center. Overall, this galaxy is consistent with an outside-in formation.

Both MRG-M0138 and MRG-M1341 have flat age gradients in the inner core. Even though the age gradients have similar patterns, Figure \ref{animes} reveals different SFHs. The SFH of MRG-M0138 remains roughly the same across the galaxy with an insignificant dip in sSFR at the center at $z\sim3$, whereas the SFH in MRG-M1341 is different in lookback times of $\sim$10 Myr to $\sim$200 Myr, i.e. the outer bins continue forming stars relative to the quenching center. Quenching in MRG-M1341 may therefore have proceeded inside out \citep[e.g.,][]{wellons2015,nelson2021}, unlike the uniform quenching in MRG-M0138. As the dominant fraction of the stellar mass in MRG-M1341 is formed much earlier in the universe (lookback times $>$1 Gyr) with almost a similar rate across the galaxy, the median age does not show a statistically significant gradient\footnote{A careful observer, however, notices that the average median-age is lower in the outskirts consistent with the SFH pattern even if it is not statistically significant.}.

MRG-M0454, MRG-M2129, MRG-P0918 and MRG-M0150 age gradients are all mildly\footnote{i.e., this non-axisymmetric trend is washed away upon azimuthally combining the data points and we only have a flat age gradient.} non-axisymmetric, with one side of the galaxy older than the other side at similar distances. This could indicate that a merger or interaction with another galaxy played a role in their formation. For three galaxies, MRG-M0150, MRG-M0454 and MRG-P0918, the SFH on one side dips below the other side for lookback times of $\sim$100 Myr to $\sim$1 Gyr, leading to an older age. The asymmetrical features are not uncommon in spatially-resolved studies of quiescent galaxies \citep[e.g.,][Figures 8 and 4, respectively]{cheng2015, setton2020}. However, this non-axisymmetric trend vanishes if we azimuthally combine the data points given their distances from the center, consistent with \citet{setton2020}.

The non-axisymmetric age gradient in MRG-M2129 is caused by the centrally concentrated 1.3 mm ALMA flux. One bin remains star-forming all along in MRG-M2129 in order to produce the significant measured 1.3 mm flux. It is interesting that this feature does not exist in the one other galaxy with a 1.3 mm ALMA detection in the REQUIEM-2D sample (MRG-M0138). This is likely because the ratio of the ALMA flux relative to other wavelengths such as $\mathrm{H_{F160W}}$ is significantly smaller than that of MRG-M2129. The ALMA detection can be roughly attributed to the 3 central bins. The fitting itself prefers the adjacent to central bin as the dominant producer of the ALMA flux in MRG-M2129. We fit another model assuming that 1.3 mm ALMA flux can only be attributed to the central bin, and the resulting SFH changed to significant star formation at the center. As the ALMA beam size is $\sim 1\farcs5 \times 1\farcs1$ \citep{whitaker2021}, we cannot distinguish between these two cases. We therefore caution that the age gradient pattern in MRG-M2129 is largely affected by the assumption for the spatial location of the 1.3 mm ALMA detection. The ALMA detection also leads to a different pattern of SFR gradients comparing to \citet{toft2017}.

With the exception of MRG-M0138\footnote{We find a super-solar metallicity for this target, also consistent with \citet{jafariyazani2020} and \citet{newman2018}.}, we measure metallicities consistent with sub-solar to solar for all of our targets, broadly in agreement with previous measurements \citep{toft2017,newman2018,jafariyazani2020,man2021}. While we do not detect any stellar metallicity gradients, there are slight hints of a relatively metal-rich center for MRG-M0454 and MRG-M1341, but our measurements are not statistically significant. 

Taken at face value, the measured flat metallicity gradient in MRG-M0138 is consistent with \citet{jafariyazani2020}. However, for the flat metallicity gradients in other targets, we note an important caveat. WFC3/G141 has a lower spectral resolution relative to the ground-based spectroscopic measurements with instruments such as Keck/MOSFIRE, it may not help with constraining potential metallicity gradients, given the estimated metallicity uncertainties here ($\sim0.3$-$0.5$ dex). These orders of magnitude uncertainties are consistent with similar studies \citep[e.g.,][]{estrada2019}. For all targets, we use the same priors for dust and metallicity for all bins, and for a half of the targets (MRG-P0918, MRG-M0150, MRG-M1341 and MRG-S1522) we get a metallicity posterior consistent with our flat prior. Therefore, our measurements cannot conclusively reject flat metallicity gradients for the REQUIEM-2D targets. Future measurements with the \emph{James Webb Space Telescope} (for example with the NIRSpec/IFU) should be suitable to answer this important question.

We find that the median stellar masses of the REQUIEM-2D galaxies (Figure \ref{fig:sfms}) are $\sim$0.1-0.5 dex higher than previously published by \citet{man2021} and \citet{newman2018} for most targets, although for some targets these measurements are consistent within the 1$\sigma$ uncertainties. This offset can be partially explained by the fact that we use non-parametric SFHs, unlike \citet{man2021} and \citet{newman2018}, and this leads to higher stellar masses and older ages \citep[e.g.,][]{leja2019}. However, the most extreme offset of $\sim$0.5 dex in MRG-M1341 is also related to an offset in the \emph{HST} photometry. We use banana-shaped apertures (Section \ref{subsec:photometry}) that efficiently enclose fluxes even for unconventional morphologies such as that of MRG-M1341 (Figure \ref{fig:color-im}), whereas \texttt{Source Extractor} \citep{bertin1996} and slit-shaped apertures are used in \citet{man2021} and \cite{newman2018}. Therefore, we measure higher fluxes for some targets in general, leading to higher stellar masses in our SED fits. Specifically, our measured magnitude for MRG-M1341 in \emph{HST} WFC3/F140W band is $\sim$0.2 dex lower (brighter).

We find $\texttt{dust\_index}<0$ for the majority of targets and $A_V<1$, consistent with the earlier studies of low-redshift galaxies \citep[e.g.,][]{salim2018}. A negative \texttt{dust\_index} corresponds to a steeper attenuation curve relative to the \citet{calzetti2000} law, leading to more UV attenuation and less IR attenuation comparably \citep{kriek2013}. We do not detect gradients in either of these dust parameters for the majority of our sample. The only exceptions are MRG-M0138 and MRG-M2129, both having a centrally concentrated 1.3 mm ALMA detection. $A_V$ is only greater than 1 magnitude for the central region of MRG-M2129 that appears to host a dust-obscured star-forming region, producing the significant observed 1.3 mm ALMA flux. It is perhaps surprising that one data point can have such a significant impact on the fit. We explore this in more detail in a future paper (Popescu et. al., in prep).

\begin{figure*}[!t]
\centering
\includegraphics[width=0.8\textwidth]{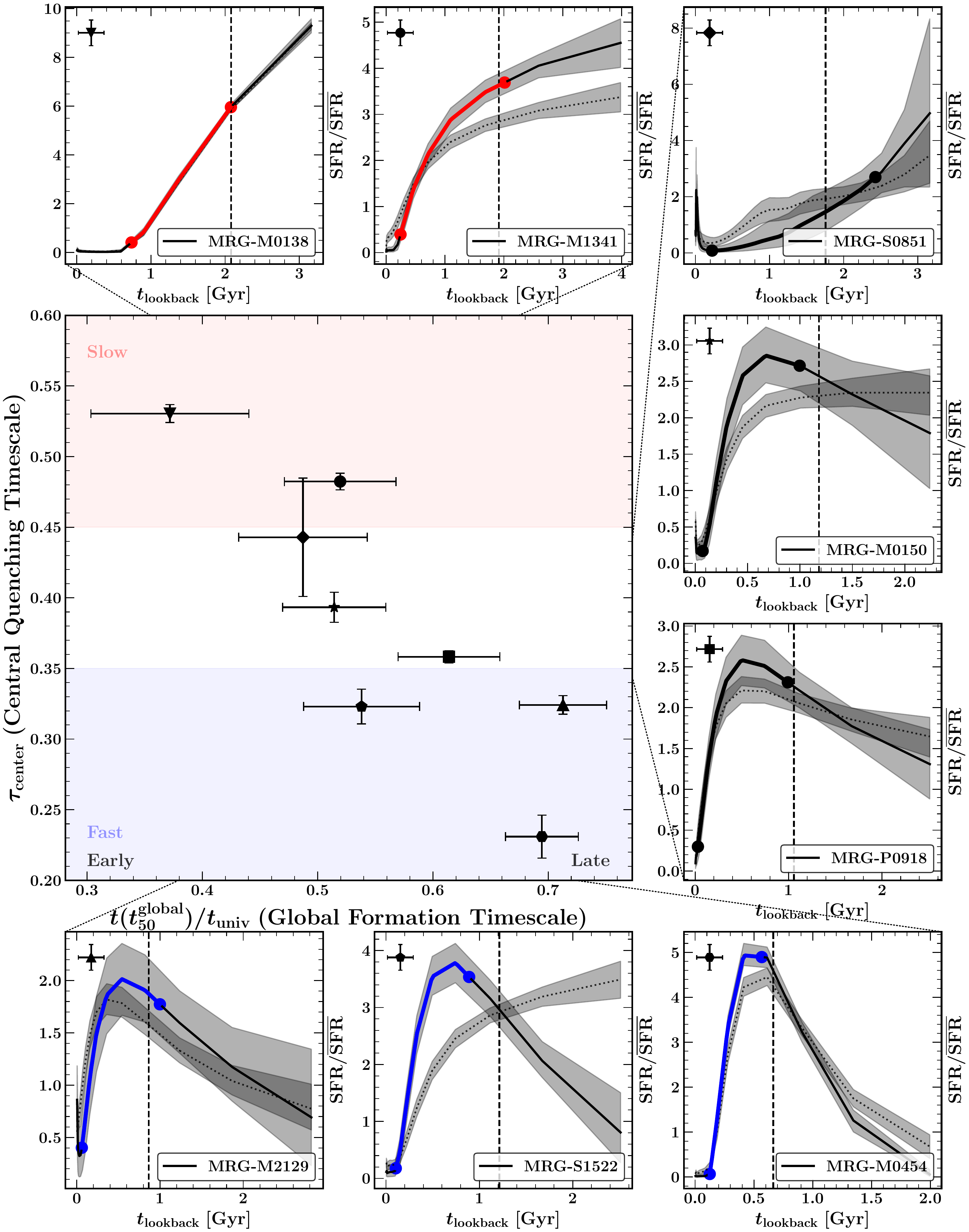}
\caption{The quenching time-scale of the central bin, defined following \citet{carnall2018}, versus the global formation time-scale defined as the ratio of the age of the universe when 50\% of the total stellar mass is formed to the age of the universe at the redshift of observation.  The global and central SFHs of all galaxies are plotted in the smaller panels. The portions of SFH that are used to define $\tau_{\mathrm{center}}$ are highlighted with blue, black and red. The vertical dashed lines in the right panels show the global measurement of $t_{50}$ (i.e., the x axis in the left panel). Quiescent galaxies that form their inner core slower (higher on the y axis, left panel) are also formed earlier in the universe (to the right on the x axis, left panel) and vice versa. \label{fig:tau-corr}}
\end{figure*} 

We are able to constrain the ages and the SFHs of the REQUIEM-2D galaxies at different physical radii thanks to the morphological diversity of the targets. The variation of the magnification-corrected physical distances across the galaxies can be used to combine the age gradients for the full sample. In Figure \ref{fig:age-grad}, we combine the measured age gradients relative to the center for the REQUIEM-2D galaxies as a function of radial distances normalized using their half-light radii, $r_c$ (see Section \ref{sec:resolved-results}, for the exact definition). We color code the age differences using the quenching time-scale, $\tau_{\mathrm{center}}$, defined following \citet{carnall2018}\footnote{$\tau = \left[t(t_{\mathrm{quench}})-t(t_{50})\right]/t(t_{\mathrm{quench}})$ where $t_{\mathrm{quench}}$ and $t_{50}$ are the lookback times when the normalized $\mathrm{SFR}/\overline{\mathrm{SFR}}=0.1$ and 50\% of the stellar mass is formed, respectively. $t(t_L)=t(z_{\mathrm{obs}})-t_L$, that is the age of the universe at lookback time of $t_L$ if the galaxy is observed at redshift of $z_{\mathrm{obs}}$.}. The age difference, $\delta t_{50}$, is defined to be the difference between the age of a given bin relative to the central bin for each target in Gyr. It is evident that most measured age differences for the REQUIEM-2D galaxies are only sampling the inner core, $r<r_c$, and are consistent with 0, i.e. a flat age gradient. However, when we consider measurements at $r\gtrsim r_c$, especially for MRG-S1522, there seems to be an upward trend. We perform two separate linear fits to the REQUIEM-2D age gradients with and without MRG-S1522. Only when including MRG-S1522 do we have a statistically significant positive slope. So while the combined age gradients for the REQUIEM-2D sample seem to be consistent with younger stellar populations in the inner core, this effect is driven by a single galaxy. To robustly confirm such a conclusion, a larger sample with more independent data points at or beyond the half-light radius is required.

We find a correlation between the quenching time-scale at the center, $\tau_{\mathrm{center}}$, and the age gradient patterns presented in Figure \ref{fig:age-grad}. The REQUIEM-2D galaxies that have smaller $\tau_{\mathrm{center}}$, that is their cores are forming fast, also have a younger stellar population in their inner core relative to their outskirts. Conversely, those galaxies that form their inner cores slowly also do so in a uniform manner, as they have flat age gradients in $r<r_c$.

To further investigate this correlation in Figure \ref{fig:age-grad}, we plot the quenching time-scale at the center versus the global formation time-scale $t(t_{50})/t_{\mathrm{univ}}$ (i.e., the ratio of the age of the universe when 50\% of the total stellar mass
formed to the age of universe at observation).  Small values for this formation timescale ratio represent galaxies that have already formed half of their stellar mass early in the universe, whereas larger values are those that form later. Interestingly, galaxies that form their cores faster (low $\tau_{\mathrm{center}}$), also seem to form later in the universe (high $t(t_{50})/t_{\mathrm{univ}}$) and vice versa. We therefore conclude that the REQUIEM-2D galaxies that are formed earlier in the universe are also forming their inner cores slowly and uniformly, whereas the galaxies in our sample that are formed later in the universe are forming their inner cores faster and have a younger center in general. The existence of fast/slow quenching channels is generally consistent with the unresolved observational studies of quiescent galaxies \citep[e.g.,][]{belli2019}, but our spatially-resolved study confirms it for the inner core and it supports that the fast quenching channel should lead to age gradients. The existence of age gradients in quiescent galaxies is also broadly consistent with cosmological simulations \citep[e.g.,][]{wellons2015, tacchella2016}.

We show that the quiescent galaxies that form later in the universe build up their cores faster and vice versa. At face value, this result is in tension with previous chemical abundance analyses that imply the opposite \citep[e.g.,][]{thomas2005}. Our observations challenge the current paradigm where the earliest massive galaxies form most rapidly. However, the chemical abundance analyses of local early-type galaxies in \citet{thomas2005} that support the current framework are unresolved and naturally probe long timescales owing to the epoch of observation. Over ten billion years, dynamical processes could in principle homogenize the central formation time-scale between $z\sim2$ and $z\sim0$ and effectively wash out the signatures that we capture at high redshift. Our observations could be consistent with a physical picture in which $z\sim2$ quiescent galaxies are formed via two different channels: central starbursts often triggered by gas-rich mergers at $z\sim2-4$ or gradual, accretion-throttled formation at early cosmic times when the Universe was denser \citet{wellons2015}. Future spatially-resolved chemical abundance studies of galaxies closer to the epoch of formation at $z\gtrsim2$ with higher spectral resolution, potentially with the \emph{James Webb Space Telescope}, are necessary to understand the origin of this tension. Our understanding of how massive galaxies formed the bulk of their stars thus may be incomplete and/or incorrect.

We finally note that the correlation between $\tau_{\mathrm{center}}$ and $t(t^{\mathrm{global}}_{50})$ does not trivially hold as the measured time-scales in principle represent different properties of the SFHs. $\tau_{\mathrm{center}}$ measures the time difference between forming 50\% of the stellar mass and when the SFR drops to 10\% of its average, $t_{\mathrm{quench}}$, normalized by $t(t_{\mathrm{quench}})$ for the center, whereas the $t(t^{\mathrm{global}}_{50})$ measures the age of the universe where 50\% of the global stellar mass is formed. For example, we have two galaxies in the sample,  MRG-M1341 and MRG-S1522, with almost the same $t(t^{\mathrm{global}}_{50})/t_{\mathrm{univ}}$ but very different quenching time-scales at their centers. As it is clear from Figure \ref{fig:tau-corr}, they also have distinct SFHs at their centers while having pretty similar global SFHs.  However, we acknowledge that a sample of eight is not adequate to sample the full parameter space possible in this diagnostic plot.  More data of similar quality would help build a more representative picture of central quenching relative to global formation timescales for the overall quiescent population at cosmic noon.

\section{Summary \label{sec:summary}}

In this paper, we present the novel analysis of deep 5-15 orbit grism spectroscopy from the REQUIEM-2D survey of eight strong-lensed quiescent galaxies ranging from $z$=1.59 to $z$=2.92.  By combining the grism spectroscopy with \emph{HST} and \emph{Spitzer} imaging, we perform a spatially resolved spectrophotometric analysis, developing and publicly releasing the \texttt{requiem2d} software package.  With this methodology, we can robustly constrain the star formation histories in the inner cores of distant galaxies in order to constrain signatures of formation and quenching.  Our conclusions from this analysis are detailed as follows. 

\begin{itemize}
    \item The REQUIEM-2D galaxies are not dusty ($A_V<1$) and have attenuation curves that are steeper than the \citet{calzetti2000} law in the UV and flatter in the IR. We cannot reject flat metallicity gradients in the REQUIEM-2D survey. 

    \item Despite having only eight targets in the REQUIEM-2D survey, we already observe a diversity of formation pathways. 
    
    \item Our measurements only include the 5 central bins (Section \ref{sec:resolved-analysis} and Figure \ref{fig:color-im}), sensitive to the inner $\sim$1-2 kpc of the galaxy cores. For these bins, corresponding to different radial distances given the morphology and gravitational magnification (see the x-axis in Figures \ref{animes}-\ref{fig:age-grad}), we have MRG-M0138 and MRG-S0851 with flat-age gradients in their inner core, consistent with an early and almost spatially-uniform formation scenario \citep{akhshik2020}. The flat age gradient in these galaxies are consistent with the results of \citet{jafariyazani2020} and \citet{setton2020}. We speculate that if there was a gradient at some point during their evolution in the inner half-light radius, it has since diminished, for example as a result of mergers. MRG-M1341 has a flat age gradient, too, but its formation pathway seems different, as the SFH implies an inside-out quenching \citep[e.g.,][]{nelson2021,tacchella2016,wellons2015}. 
    
    \item We detect a statistically significant age gradient in MRG-S1522, with the galaxy looking younger in the inner half-light radius. This result is consistent with \citet{gobat2017}, and it may imply non-uniform quenching \citep{wellons2015}. The recovered SFH of MRG-S1522 clearly indicates that this galaxy quenches in the outer bins first. 
    
    \item For the remaining four REQUIEM-2D galaxies, we find non-axisymmetric age gradients. Similar asymmetries are reported for both local and high redshift galaxies \citep[e.g.,][]{cheng2015,setton2020}.  While this asymmetry may have significant implications for quenching, we note that by combining data points azimuthally, it washes out any statistically significant gradients. We also note that some massive rotationally-supported star-forming galaxies at $z\sim 2$ show slight disturbances in their morphology. \citep{girard2018}.
    
    \item While the combined age gradients for the REQUIEM-2D sample are consistent with younger stellar populations in the inner half-light radius, only one galaxy (MRG-S1522) samples the age beyond the half-light radius.  A larger sample with more independent data points at or beyond the half-light radius are required to reach a robust conclusion. 
    
    \item By investigating the age gradient patterns, the global formation time-scales, and quenching time-scales at the center, we conclude that the REQUIEM-2D galaxies that are forming their inner cores slowly and uniformly are consistent with an early formation scenario, whereas the galaxies that are forming their inner half-light radius faster are younger at center and are generally formed later in the universe.
    
    \item Our main result is in tension with previous unresolved chemical abundance analyses of local early-type galaxies \citep[e.g.,][]{thomas2005}, supporting the current paradigm where the earliest massive galaxies form most rapidly. While dynamical processes can alleviate this tension, future spatially-resolved chemical abundance studies closer to the epoch of formation with higher spectral resolution are necessary to understand its origin.

\end{itemize}

Future high spatial and spectral resolution observations of lensed quiescent galaxies, such as those possible with the \emph{James Webb Space Telescope}, can provide further insights to the metallicity, age and chemical abundance gradients in the quiescent population to better constrain their formation pathways. For example, using NIRSpec/IFU, one should be able to constrain metallicity gradients due to NIRSpec's superior spectral resolution relative to the G141 grism spectroscopy used herein. Also, by using the IFU, one can avoid the complications of the contamination by nearby objects in grism spectroscopy. This analysis serves as a jumping off point, proposing a new diagnostic of formation pathways (i.e., Figure~\ref{fig:tau-corr}) that can be tested with larger samples, requiring high spatial resolution and enough signal-to-noise in the outskirt of galaxies.

\section*{acknowledgements}
M.A. gratefully acknowledges support by NASA under award No 80NSSC19K1418. This work is based on observations made with the NASA/ESA Hubble Space Telescope, obtained at the Space Telescope Science Institute, which is operated by the Association of Universities for Research in Astronomy, Inc., under NASA contract NAS 5-26555. M.A gratefully acknowledges support from STScI grants GO-14622 and GO-15633. K.W. wishes to acknowledge funding from the Alfred P. Sloan Foundation. S.T. and G.B. acknowledge support from the ERC Consolidator Grant funding scheme (project ConTExt, grant number No. 648179). The Cosmic Dawn Center is funded by the Danish National Research Foundation under grant No. 140. The Dunlap Institute is funded through an endowment established by the David Dunlap family and the University of Toronto. G.M. acknowledges funding from the European Union’s Horizon 2020 research and innovation program under the Marie Skłodowska-Curie grant agreement No MARACHAS - DLV-896778. H.E. gratefully acknowledges support from STScI grant GO-15466. C.C.W. acknowledges support from NIRCam Development Contract NAS5-02105 from NASA Goddard Space Flight Center to the University of Arizona.

\bibliography{sample62}

\appendix

\section{Gravitational Lensing model \label{sec:lensing}}

MRG-P0918 is a highly magnified, singly-imaged galaxy lensed by the massive galaxy cluster PSZ1 G295.24-21.55 at $z=0.61$ \citep{planck2014}. We make use of the \emph{HST} images to locate three multiple lensed systems (numbered 1 to 3), composed of 3 images each. We present these multiple images in Table \ref{tab:mul0918}, adopting  the usual notation X.Y where X identifies a system of multiple images and Y is a running number identifying all images for a given system. System 1 has a known spectroscopic redshift with $z=2.146$. This redshift is measured using \texttt{Grizli} and its automated redshift fitting pipeline leveraging grism spectroscopy (observed as a part of HST-GO-15663). In particular, the [O III] doublet at rest-frame wavelengths of 4960/5008 $\mathrm{\AA}$ is used to constrain the redshift with a reduced $\chi^2$ of 1.04. However, systems 2 and 3 have an unknown redshift.

We model the mass distribution of PSZ1 G295.24-21.55 using the latest version (v7.1) of the public {\sc Lenstool} software\footnote{Publicly available at:\\ \url{https://git-cral.univ-lyon1.fr/lenstool/lenstool}}\citep{jullo2007}. Following previous work on cluster lens models \citep[e.g.,][]{Richard2014}, we assume the total mass distribution to be a combination of double Pseudo Isothermal Elliptical mass profiles parametrised with a velocity dispersion $\sigma^*$, a core radius $r_{\rm core}$ and a cut radius $r_{\rm cut}$ \citep{Suyu2010}. The HST images show that the cluster is dominated by a central Brightest Cluster Galaxy (BCG), and we therefore model the smooth cluster-scale mass distribution using a single dPIE  mass distribution. Galaxy-scale mass components are added as individual dPIE profiles, where we select galaxies based on the cluster red sequence from \emph{HST} color F555W-F814W. To reduce the number of parameters following previous works \citep[e.g.,][]{Richard2010}, we assume that the mass component parameters associated with the cluster members follow a scaling relation with respect to $\sigma^*$ and $r_{\rm cut}^*$ for a $L^*$ galaxy. We fix $r_{\rm cut}^*=45$ kpc to remove degeneracy between these parameters. This does not affect the predictions from the lens model for MRG-P0918.

\textsc{Lenstool} uses a  Monte Carlo Markov Chain (MCMC) to sample the posterior probability distribution of the model, expressed as a function of the likelihood defined in \citet{jullo2007}. In practice, we minimise the distances in the image plane:

\begin{equation}
\chi^{2} = \sum\limits_{i,j}  \frac{ {\Vec{\theta^{(i,j)}_{obs}} - \Vec{{\theta^{(i,j)}_{pred}}}}^{2}}{\sigma_{pos}^{2}}\ ,
\end{equation}

\noindent with $\Vec{\theta^{(i,j)}_{obs}}$ and $\Vec{\theta^{(i,j)}_{pred}}$ representing the observed and predicted vector positions of the multiple image $j$ in system $i$, respectively. Furthermore, $\sigma_{pos}$ is a global error on the position of all multiple images, which we fix at 0\farcs5.

Using all three systems from Table \ref{tab:mul0918} as constraints, we obtain a best fit rms of $1.4\arcsec$ for this mass model. The best fit parameters for the mass distribution and the optimised redshifts of systems 2 and 3 are presented in Tables \ref{tab:model0918} and \ref{tab:mul0918} respectively. Based on the best fit mass model of the cluster, we predict a magnification of $\mu=7.69\pm0.86$ for MRG-P0918. We also summarize all the measured magnifications and half-light radii in Table \ref{tab:murc}.

\begin{table*}
    \centering
\begin{tabular}{lllll}[!ht]
ID & $\alpha$ & $\delta$ & $z_{\rm prior}$ & $z_{\rm model}$ \\
   & [deg] & [deg] & & \\
\hline 
 1.1 & 09:18:51.68 & -81:03:18.50 & 2.146 & \\
 1.2 & 09:18:49.92 & -81:03:10.51 & 2.146 & \\
 1.3 & 09:18:50.60 & -81:02:32.93 & 2.146 & \\
 2.1 & 09:19:00.56 & -81:03:10.22 & [0.8-5.0] & 1.66$\pm$0.04\\
 2.2 & 09:18:59.95 & -81:03:15.40 &  & \\
 2.3 & 09:18:58.98 & -81:02:43.44 &  & \\
 3.1 & 09:18:49.70 & -81:03:28.89 & [0.8-5.0] & 1.87$\pm$0.05\\
 3.2 & 09:18:45.48 & -81:03:01.81 &  & \\
 3.3 & 09:18:47.26 & -81:02:47.12 &  & \\
\hline
\end{tabular}
\caption{\label{tab:mul0918}List of multiply imaged systems used as constraints in the MRG-P0918 mass model. From left to right: identification number for the multiple image, sky coordinates, redshift constraints from spectroscopy. Redshifts with error bars are not constrained with spectroscopy and are predictions from our lens model.}
\end{table*}

\begin{table*}
    \centering
\begin{tabular}{llllllll}[!t]
Potential & $\Delta\alpha$ & $\Delta\delta$ & $e$ & $\theta$ & r$_{\rm core}$ & r$_{\rm cut}$ & $\sigma$ \\
   & [arcsec] & [arcsec] & & [deg] & kpc & kpc & km/s \\
\hline 
DM & $ -5.0^{+  0.9}_{ -0.0}$ & $ -0.5^{+  0.5}_{ -0.3}$ & $ 0.50^{+ 0.00}_{-0.01}$ & $176.9^{+  0.6}_{ -0.4}$ & $140^{+18}_{-10}$ & $[1000]$ & $1319^{+43}_{-25}$ \\
L$^{*}$ galaxy &  & & & & $[0.15]$ & $[45]$ & $127^{+7}_{-17}$\\
\hline
\end{tabular}

    \caption{\label{tab:model0918}Mass model parameters for MRG-P0918. The reference world coordinate system (WCS) location corresponds to the brightest cluster member (09:18:51.11,$-$81:03:04.35). From left to right: mass component, position relative to the cluster centre ($\Delta$R.A. and $\Delta$Dec.), dPIE shape (ellipticity and orientation), core and cut radii and velocity dispersion. The final row  is the generic galaxy mass at the characteristic luminosity L$^*$, which is scaled to match each of the cluster member galaxies.  Parameters in square brackets are fixed {\it a priori} in the model.
    }
\end{table*}

\begin{table}
    \centering
\begin{tabular}{|c|c|c|}
\hline
Target & $\mu$ & $r_c$ [Arcsec] \\ \hline \hline
MRG-M1341 & 30$\pm$8 & 0\farcs32$\pm$0\farcs04 \\ \hline
MRG-M0138 & 13$\pm$5 & 0\farcs4$\pm$0\farcs1 \\ \hline
MRG-M0150 & 5$\pm$1 & 0\farcs2$\pm$0\farcs03 \\ \hline
MRG-P0918 & 7.7$\pm$0.9 & 0\farcs22$\pm$0\farcs01 \\ \hline
MRG-S1522 & 15$\pm$6 & 0\farcs06$\pm$0\farcs01 \\ \hline
MRG-M2129 & 4.6$\pm$0.2 & 0\farcs25$\pm$0\farcs01 \\ \hline
MRG-M0454 & 11$\pm$2 & 0\farcs13$\pm$0\farcs01 \\ \hline
\end{tabular}
\caption{The values of the linear gravitational magnification $\mu$, and the source plane half-light radius $r_c$ defined in Section \ref{sec:resolved-results}. The gravitational magnifications are from this appendix and \citet{ebeling2018,  man2021, newman2018, rodney2021, sharon2020}. \label{tab:murc}}

\end{table}

\section{Joint spectrophotometric fit \label{sec:plots}}
\begin{figure*}[!th]
\gridline{\fig{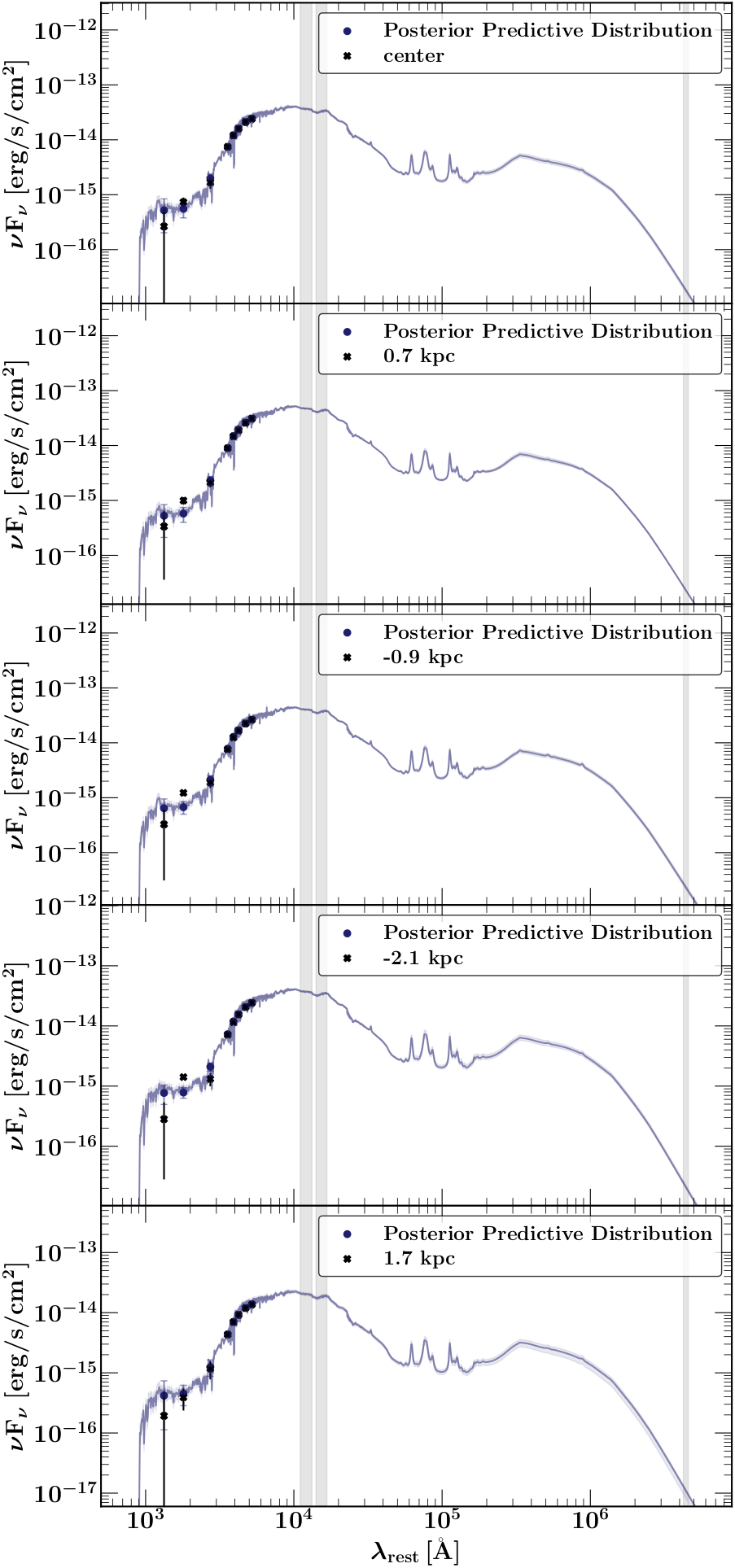}{0.45\textwidth}{}
\hspace{0in}\fig{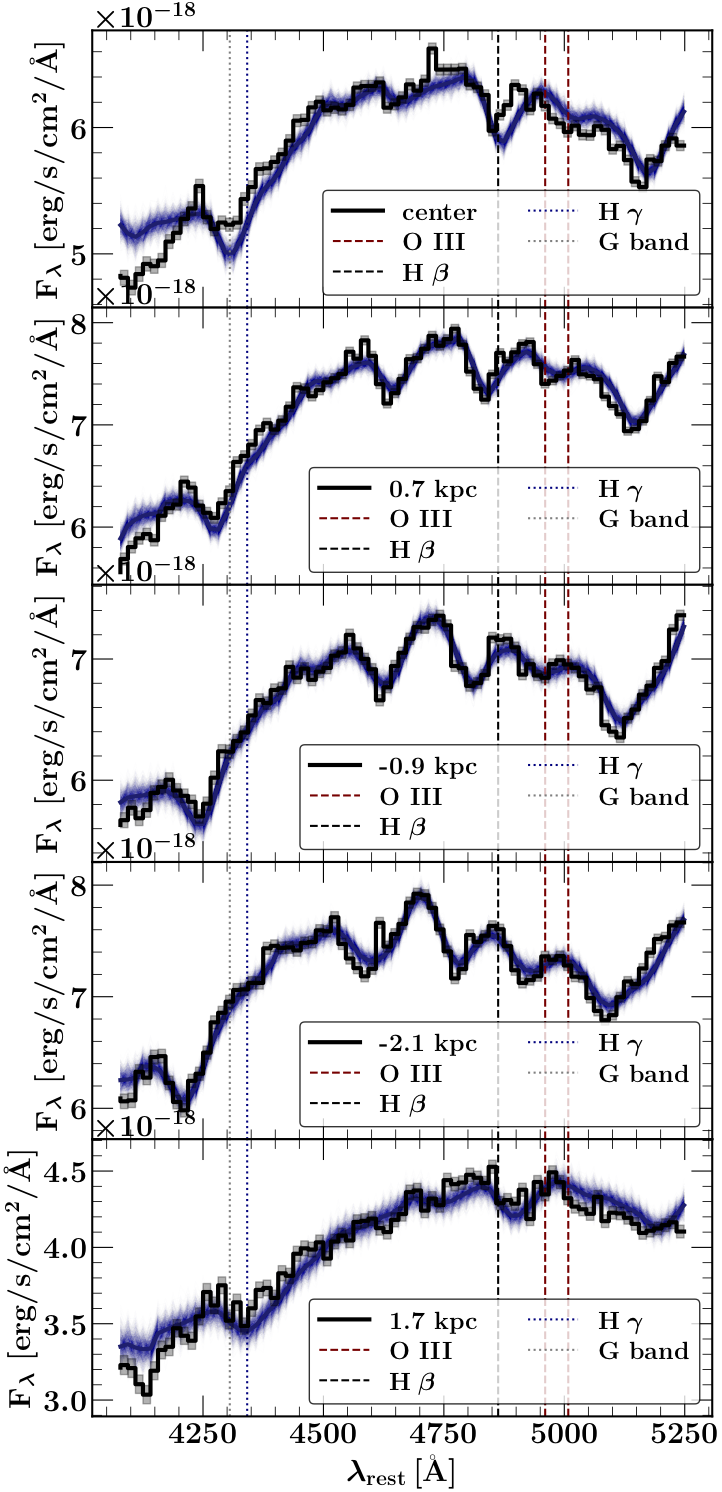}{0.47\textwidth}{}}
\vspace{-0.3in}
\caption{\textbf{Left:} the SED of MRG-M0138. The blue line shows 200 draws from the FSPS model. The shaded regions in the SED coincide with the unresolved \emph{Spitzer} IRAC channels 1 and 2 and semi-resolved ALMA 1.3 mm band passes. \textbf{Right:} The resolved extracted 1D grism spectra. \label{fig:resolved-mrg-m0138}}
\end{figure*}

\begin{figure*}[!th]
\gridline{\fig{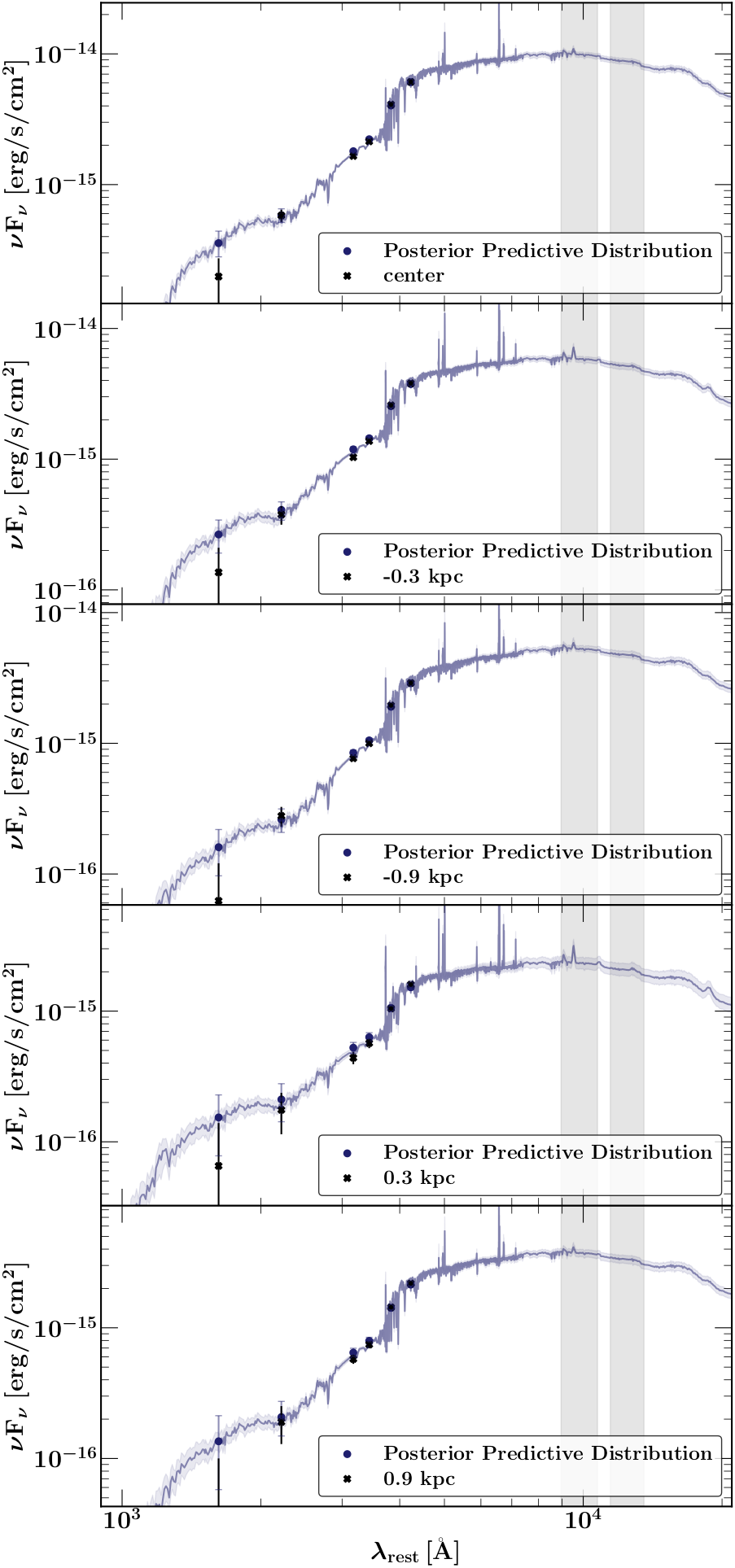}{0.45\textwidth}{}
\hspace{0in}\fig{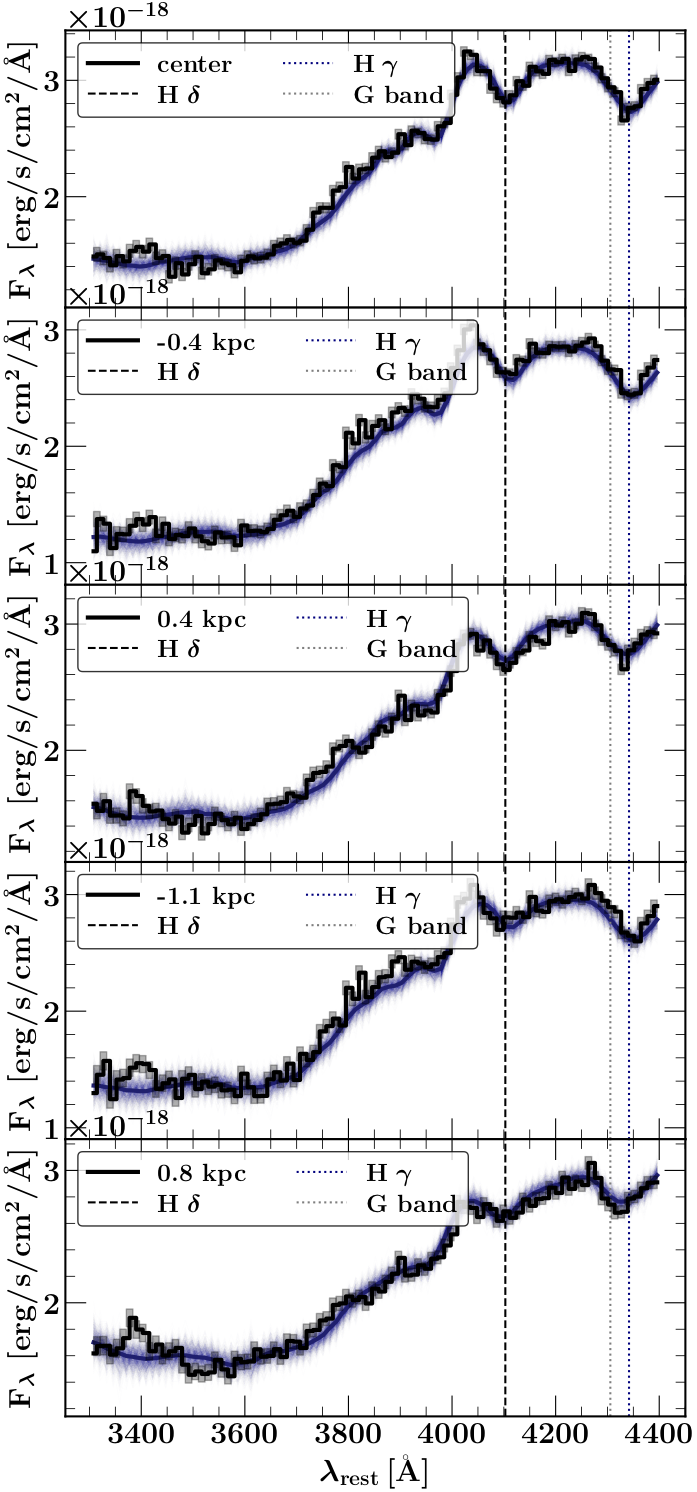}{0.455\textwidth}{}}
\vspace{-0.3in}
\caption{\textbf{Left:} The SED of MRG-M0150. The blue line shows 200 draws from the FSPS model. The shaded regions in the SED coincides with the unresolved \emph{Spitzer} IRAC channels 1 and 2 band passes. \textbf{Rihgt:} The resolved extracted 1D grism spectra. \label{fig:resolved-mrg-m0150}}
\end{figure*}

\begin{figure*}[!th]
\gridline{\fig{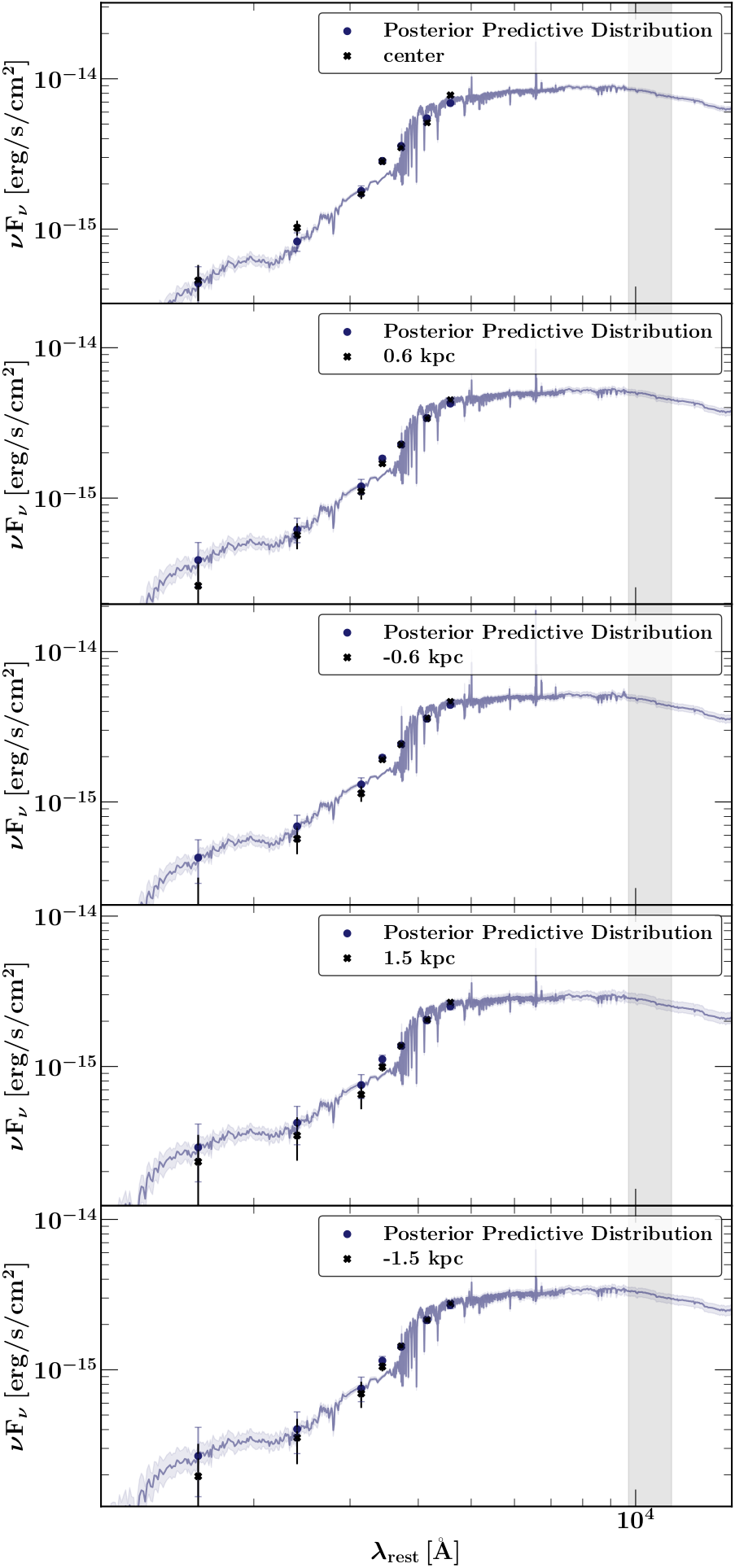}{0.45\textwidth}{}
\hspace{0in}\fig{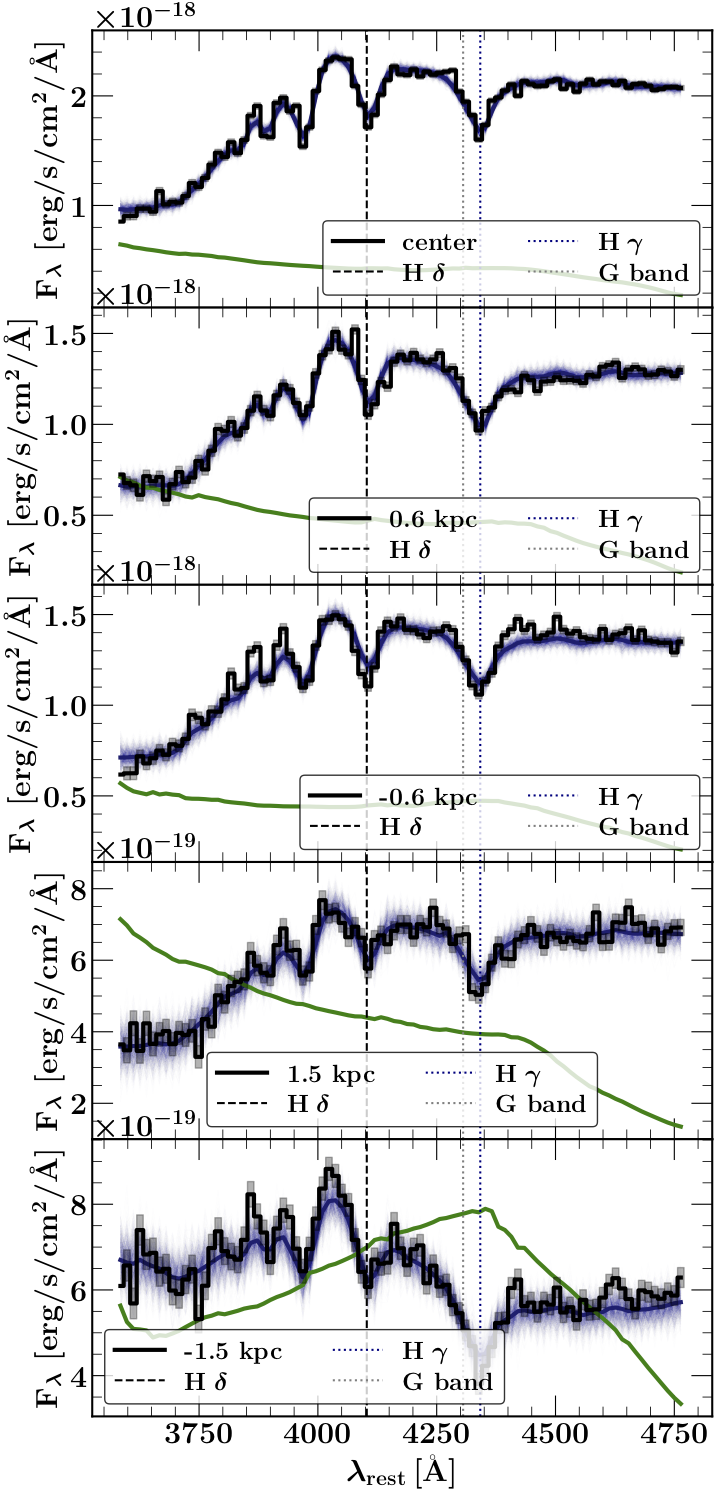}{0.465\textwidth}{}}
\vspace{-0.3in}
\caption{\textbf{Left:} The SED of MRG-P0918. The blue line shows 200 draws from the FSPS model. The shaded regions in the SED coincides with the unresolved \emph{Spitzer} IRAC channel 1 band pass. \textbf{Right:} The resolved extracted 1D grism spectra. The green line in the spectra denotes the collapsed 1D contamination, affecting the shape of the spectra of the outer bins significantly. \label{fig:resolved-mrg-p0918}}
\end{figure*}

\begin{figure*}[!th]
\gridline{\fig{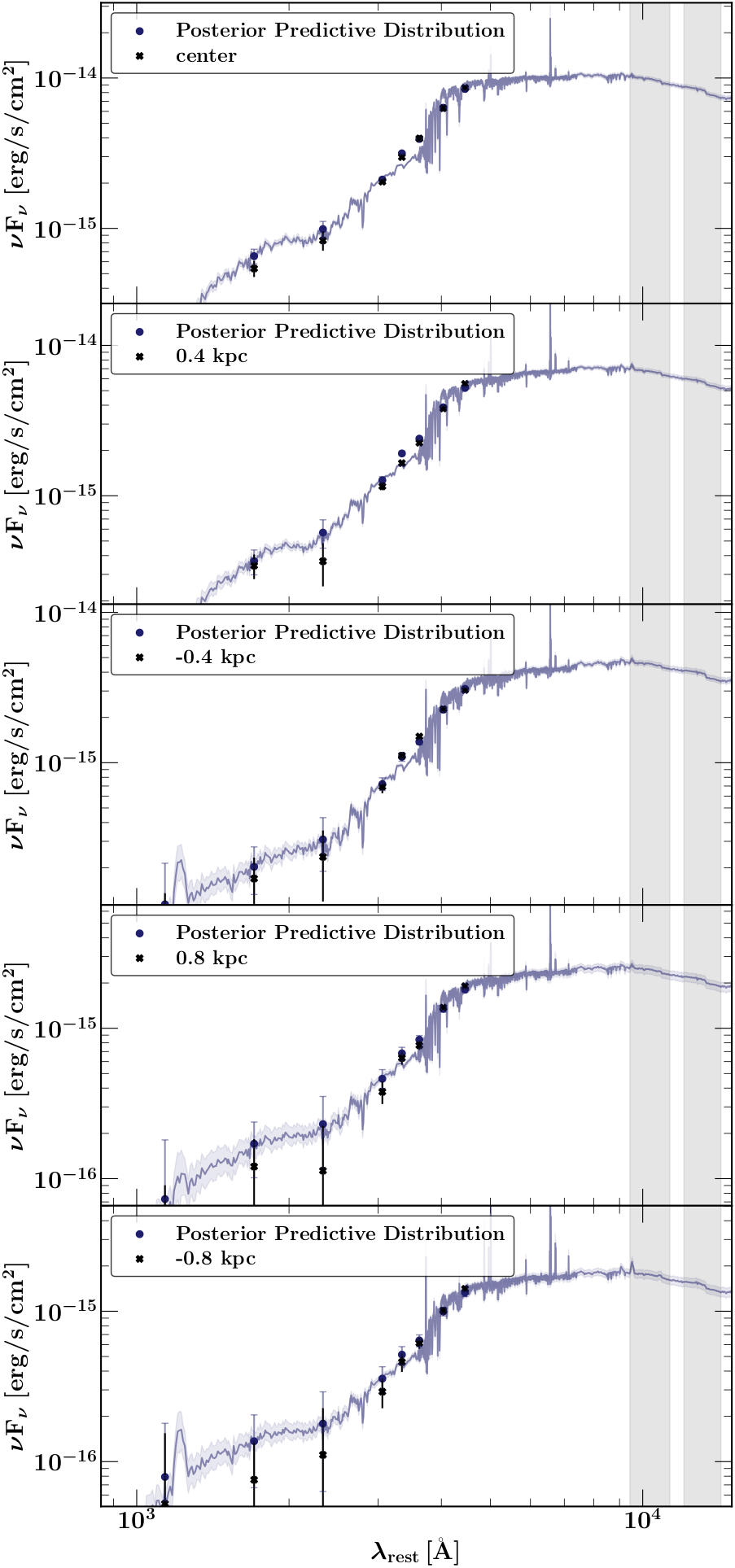}{0.45\textwidth}{}
\hspace{0in}\fig{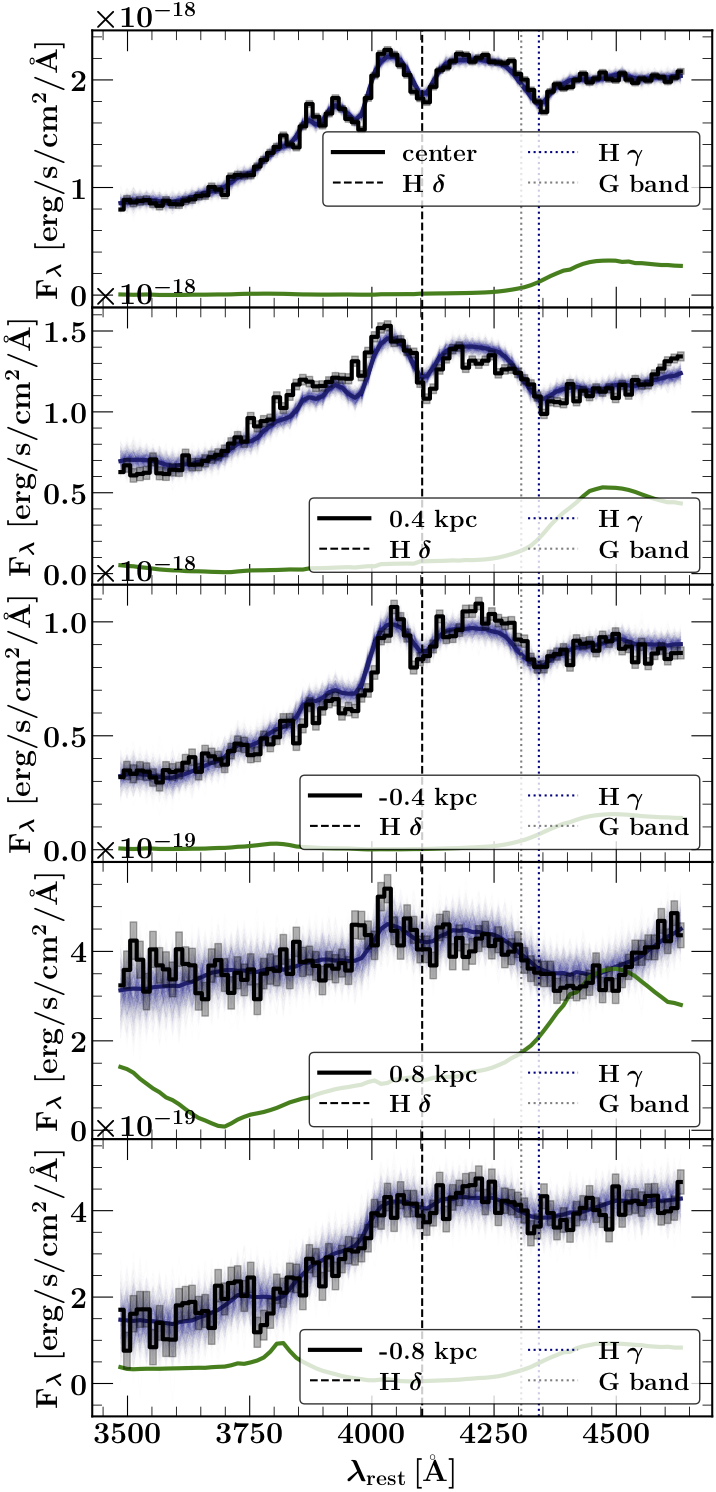}{0.47\textwidth}{}}
\vspace{-0.3in}
\caption{\textbf{Left:} The SED of MRG-S1522. The blue line shows 200 draws from the FSPS model. The shaded regions in the SED coincides with the unresolved  \emph{Spitzer} IRAC channels 1 and 2 band passes. \textbf{Right:} The resolved extracted 1D grism spectra The green line in the spectra denotes the collapsed 1D contamination, affecting the shape of the spectra of the outer bins significantly. \label{fig:resolved-mrg-s1522}}
\end{figure*}

\begin{figure*}[!th]
\gridline{\fig{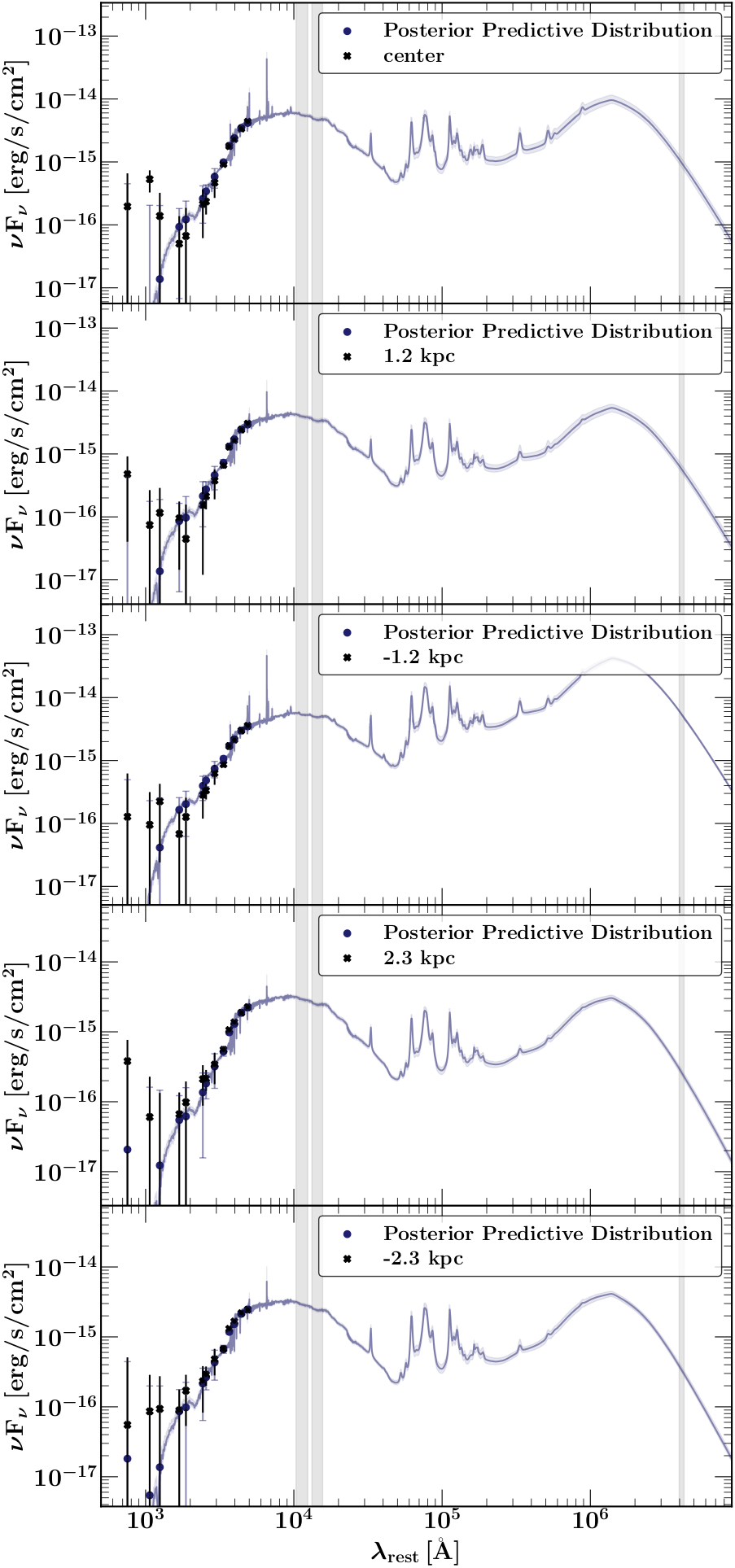}{0.45\textwidth}{}
\hspace{0in}\fig{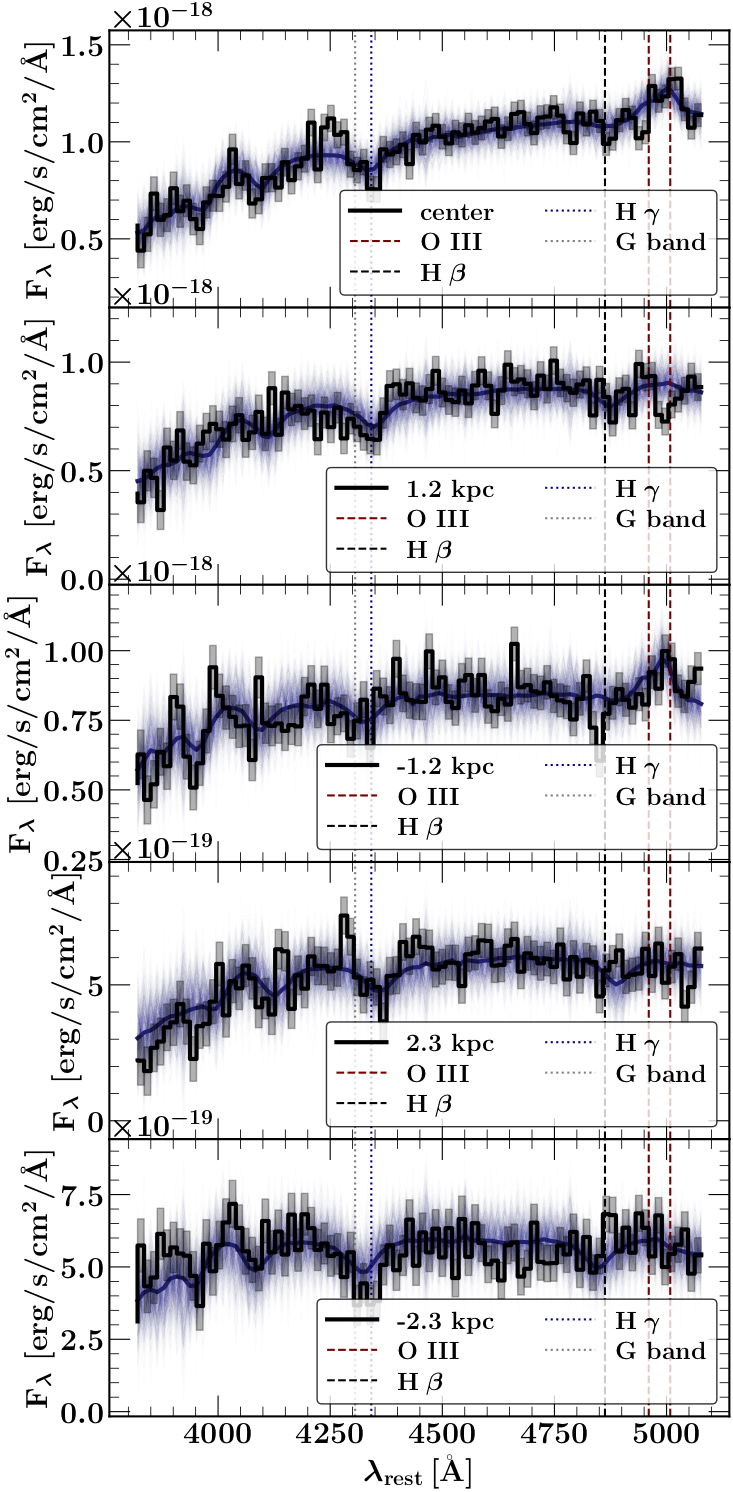}{0.48\textwidth}{}}
\vspace{-0.3in}
\caption{\textbf{Left:} The SED of MRG-M2129. The blue line shows 200 draws from the FSPS model. The shaded regions in the SED coincides with the unresolved \emph{Spitzer} IRAC channels 1 and 2 as well as ALMA 1.3mm band passes. \textbf{Right:} The resolved extracted 1D grism spectra. \label{fig:resolved-mrg-m2129}}
\end{figure*}

\begin{figure*}[!th]
\gridline{\fig{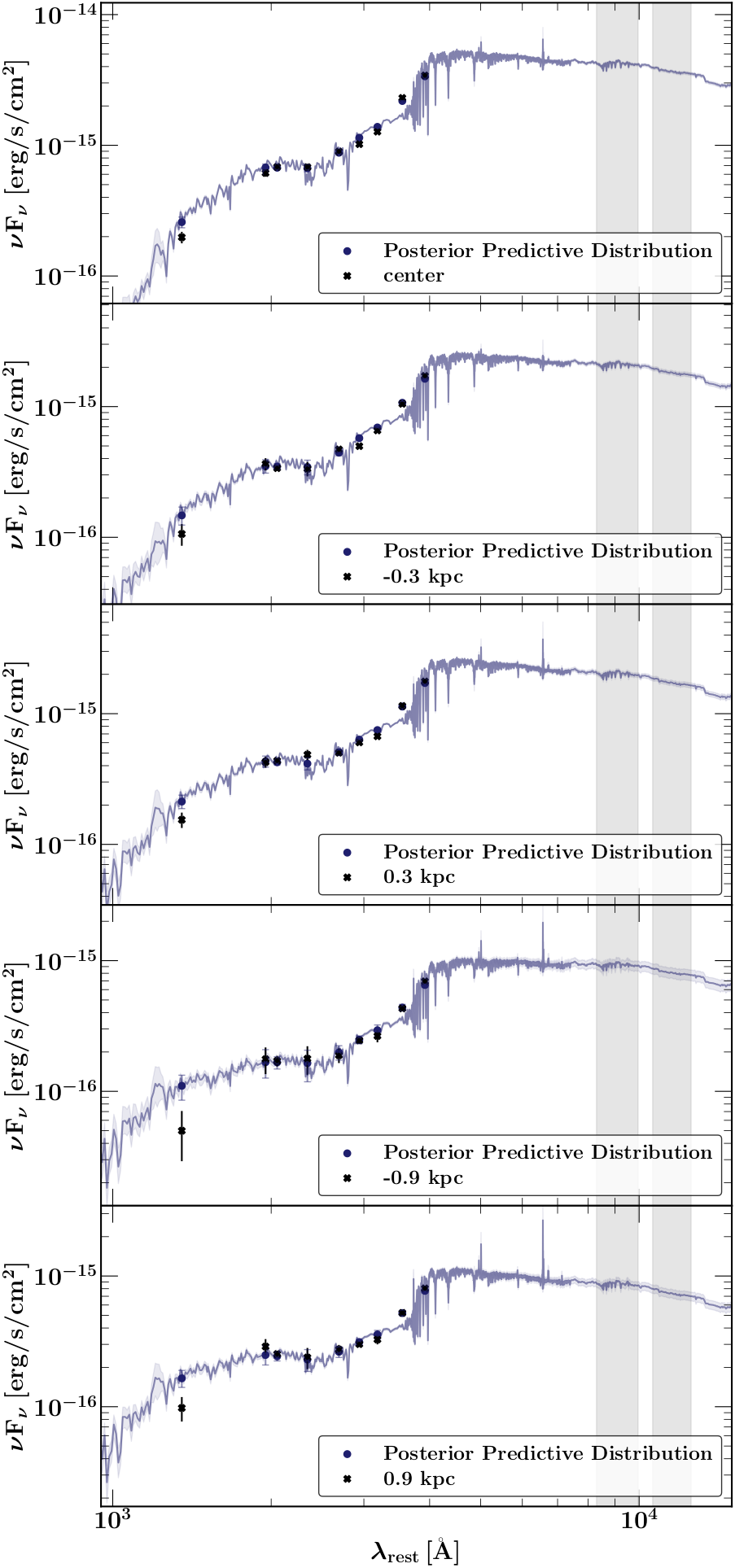}{0.45\textwidth}{}
\hspace{0in}\fig{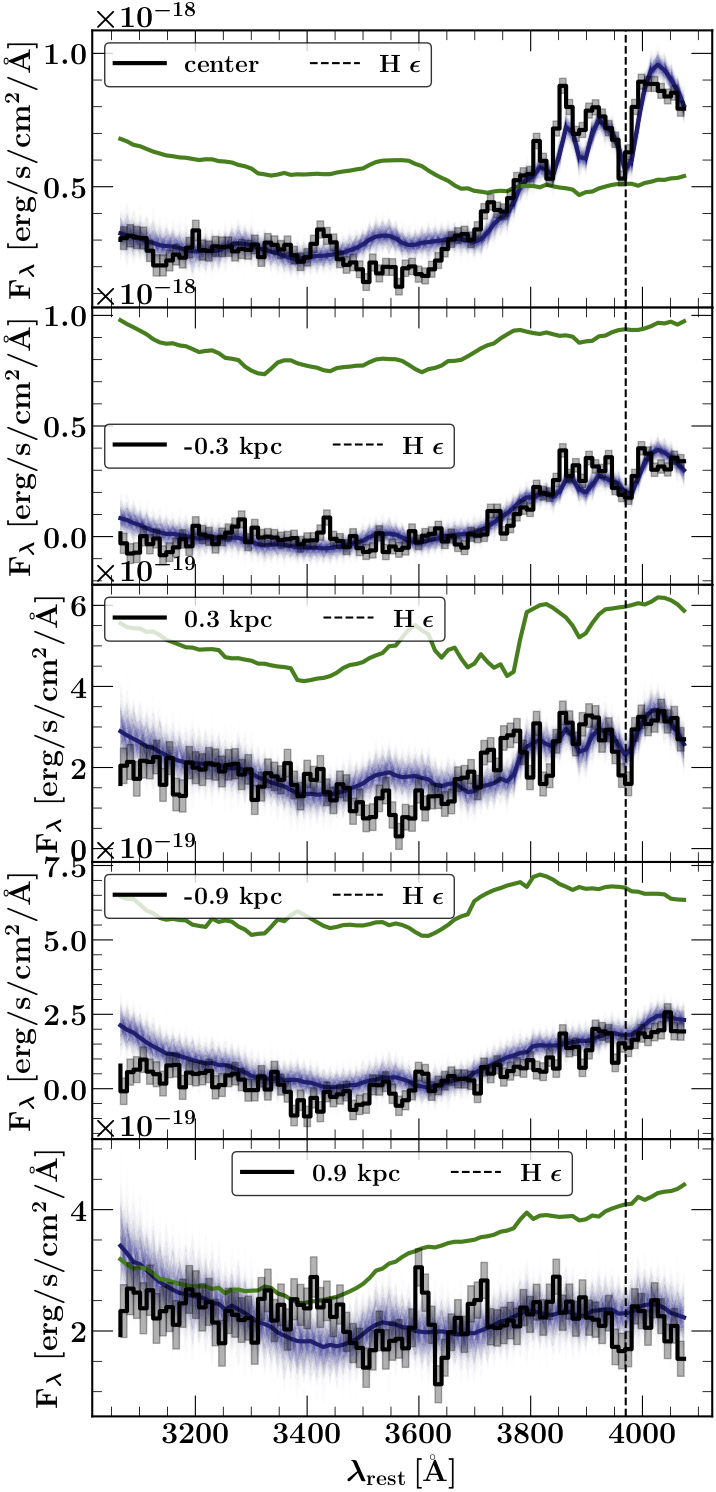}{0.475\textwidth}{}}
\vspace{-0.3in}
\caption{\textbf{Left:} The SED of MRG-M0454. The blue line shows 200 draws from the FSPS model. The shaded regions in the SED coincides with the unresolved \emph{Spitzer} IRAC channels 1 and 2. \textbf{Right:} The resolved extracted 1D grism spectra. The green line in the spectra denotes the collapsed 1D contamination, affecting the shape of the spectra of all bins significantly.  \label{fig:resolved-mrg-m0454}}
\end{figure*}

We discussed the joint spectrophotmetric fit in Section \ref{sec:resolved-results} and present the results for one of the targets, MRG-M1341, there (Figure \ref{fig:resolved-mrg-m1341}). We show the similar plots for the rest of the targets here in Figures \ref{fig:resolved-mrg-m0138}-\ref{fig:resolved-mrg-m0454}.

Despite a careful selection of the grism dispersion angles, it is impossible to fully remove the contamination from the outer fainter wings of these galaxies because the targets are dispersed perpendicular to their longer axis of symmetry and are also located in the crowded fields of views (lensed by clusters). Therefore, the contamination of nearby sources in WFC3/G141 grism data remains significant for three targets: MRG-P0918, MRG-S1522 and MRG-M0454. We show the collapsed 1D spectrum of the contamination in green. The contamination is modeled and removed before fitting following \citet{akhshik2020}, but this procedure is not perfect and therefore the shape of the 1D collapsed spectra is affected to some extent for the mentioned targets (Figures \ref{fig:resolved-mrg-p0918}, \ref{fig:resolved-mrg-s1522} and \ref{fig:resolved-mrg-m0454}, right panels). We finally note that the full \texttt{requiem2d} model is constructed and fit in the native 2D space of the WFC3/G141 spectra whereas the 1D collapsed spectra are constructed only for visual purposes. 

\end{document}